\documentstyle[aps,epsf,prd,floats,cite,preprint]{revtex}

\def\lsim{\mathrel{\raise.3ex\hbox{$<$\kern-.75em\lower1ex\hbox{$\sim$}}}}
\def\gsim{\mathrel{\raise.3ex\hbox{$>$\kern-.75em\lower1ex\hbox{$\sim$}}}}

\newcommand{\nn}{\nonumber}

\newcommand{\bmp}{\mbox{\boldmath $p$}}
\newcommand{\bmq}{\mbox{\boldmath $q$}}
\newcommand{\bmk}{\mbox{\boldmath $k$}}

\newcommand{\bilin}[3]{{\bar #1\, #2 \, #3}}
\newcommand{\me}[3]{{\left\langle{#1}\vphantom{#2 #3} 
  \right|{#2}\left|\vphantom{#1 #2}{#3}\right\rangle}}

\newcommand{\vslash}{\raise-.15ex\hbox{/}\mkern-9.5mu v}
\newcommand{\qslash}{\raise-.10ex\hbox{/}\mkern-8.5mu q}

\newcommand{\pslash}{\raise-.15ex\hbox{/}\mkern-9.5mu p}

\newcommand{\SS}{\mbox{1S}}
\newcommand{\sSS}{\mbox{\scriptsize 1S}}

\def\O#1#2{\mbox{\boldmath $O$}_{\mbox{\scriptsize\boldmath $#1$},#2}}

\def\Od#1#2{\mbox{\boldmath $O$}^\dagger_{\mbox{\scriptsize\boldmath $#1$},#2}}
\def\itbf#1{\mbox{\boldmath $#1$}}
\def\sitbf#1{\mbox{\scriptsize\boldmath $#1$}}

\def\bsigma{\mbox{\boldmath $\sigma$}}
\def\bgamma{\mbox{\boldmath $\gamma$}}
\def\lqcd{\Lambda_{\rm QCD}}
\def\ms{$\overline{\rm MS}$ }
\def\psip#1{\psi_{\mbox{\scriptsize\boldmath $#1$}}}
\def\chip#1{\chi_{\mbox{\scriptsize\boldmath $#1$}}}
\def\abs#1{\left| #1 \right|}
\def\OMIT#1{}

\begin{document}

\preprint{\vbox{\tighten \hbox{MPI-PhT/2001-17} \hbox{UCSD/PTH 2001--06} 
                 \hbox{PITHA 01/07}
                 \hbox{} } }
\title{ The Threshold $t\bar t$ Cross Section at NNLL Order  }

\author{ A.~H.~Hoang${}^a$, A.~V.~Manohar${}^b$,
         I.~W.~Stewart${}^b$, and T.~Teubner${}^c$\\[3mm]}

\address{${}^a$ Max-Planck-Institut f\"ur Physik (Werner-Heisenberg-Institut),
  \\[-3mm] F\"ohringer Ring 6, 80805 M\"unchen, Germany}
\address{${}^b$ Department of Physics, University of California at San Diego,
  \\[-3mm] 9500 Gilman Drive, La Jolla, CA 92093-0319, USA}
\address{${}^c$ Institut f\"ur Theoretische Physik E, RWTH Aachen,
   D-52056 Aachen, Germany}

\maketitle

{\tighten
\begin{abstract}

The total cross section for top quark pair production close to threshold in
$e^+e^-$ annihilation is investigated. Details are given about the calculation
at next-to-next-to-leading logarithmic order.  The summation of logarithms leads
to a convergent expansion for the normalization of the cross section, and small
residual dependence on the subtraction parameter $\nu$. A detailed analysis of
the residual $\nu$ dependence is carried out. A conservative estimate for the
remaining uncertainty in the normalization of the total cross section from QCD
effects is $\delta\sigma_{t\bar t}/\sigma_{t\bar t}\lesssim \pm 3\%$. This makes
precise extractions of the strong coupling and top width feasible, and further
studies of electroweak effects mandatory.

\end{abstract}
}

\tighten

\section{Introduction}
\label{sectionintroduction}

Detailed studies of the top quark are among the major projects of a future
lepton pair collider. Of particular interest is the production of top quark
pairs close to threshold, where the top quark velocity $v$ is small and the
usual perturbative expansion in terms of the strong coupling breaks down due to
Coulomb singularities. The large top quark width $\Gamma_t\sim 1.5$~GeV
prohibits the production of toponium states and leads to a cross section that
rises smoothly when top pair production becomes kinematically
allowed. Furthermore, the large decay rate, $\Gamma_t\gg\Lambda_{\rm QCD}$,
serves as an infrared cutoff, allowing the use of perturbative methods for the
description of the non-relativistic top-antitop dynamics to a high degree of
precision\ \cite{Kuehn1,Fadin1}.

In the past numerous studies have been carried out to assess the feasibility and
precision for extracting various top quark properties from a threshold run\
\cite{Strassler1,Jezabek1,Sumino1,Murayama1,Harlander1,Fuji1,Comas1}.  Recently,
next-to-next-to-leading order (NNLO) QCD corrections to the total cross section
were calculated using the concept of effective field theories\
\cite{Hoang1,Melnikov1,Yakovlev1,Beneke1,Nagano1,Hoang2,Penin1}. Surprisingly,
the corrections were found to be as large as the next-to-leading order (NLO) QCD
corrections. From the residual scale dependence in the NNLO result, the
normalization of the cross section was estimated to have $\approx 20\%$
uncertainty~\cite{Hoang3}. It was concluded that top quark short-distance mass
parameters can be determined with a precision of $200$~MeV or better from the
shape of the cross section, if so-called threshold mass parameters are employed\
\cite{Hoang3,Peralta1}. However, the large NNLO QCD corrections to the
normalization of the cross section seemed to jeopardize competitive measurements
of the top width, strong top coupling\ \cite{Peralta1}, or the top Yukawa
coupling in the case of a light Higgs. Moreover, the large NNLO corrections
seemed to indicate that, despite the perturbative nature of the $t\bar t$
system, high precision computations are not feasible.

A common feature of all the NNLO QCD calculations is that the running from the
hard scale $m_t$ down to the non-relativistic scales which govern the dynamics
of the top-antitop system was not taken into account. In other words, at NNLO
potentially large QCD logarithms of ratios of the hard scale and the
non-relativistic scales $m_t v$ and $m_t v^2$ were treated perturbatively. From
weak decays of $K$ and $B$ mesons, it is known that a consistent summation of
logarithms between $m_W$ and the low-energy hadronic scales can significantly
change the magnitude of Wilson coefficients and the corresponding physical
predictions (see for e.g. Ref.~\cite{Kphysics2}). In the case of the top quarks
near threshold all the scales are well separated, $m_t\gg m_t v\gg m_t
v^2\gg\Lambda_{\rm QCD}$. Since $\ln[m_t/(m_t v^2)]\sim \ln(40)=3.7$, these
logarithmic contributions can be sizeable (see for e.g. Ref.~\cite{Kniehl1}).

In this work details are presented for the computation in Ref.~\cite{Hoang8}
where the impact of the summation of QCD logarithms of $v$ on the photon induced
total cross section were examined. We also add the vector and axial-vector
contributions from $Z$ exchange. In this framework the expansion for the
normalized cross section $R$ takes the form
\begin{eqnarray}
 R \, = \, \frac{\sigma_{t\bar t}}{\sigma_{\mu^+\mu^-}}
 \, = \,
 v\,\sum\limits_k \left(\frac{\alpha_s}{v}\right)^k
 \sum\limits_i \left(\alpha_s\ln v \right)^i \times
 \bigg\{1\,\mbox{(LL)}; \alpha_s, v\,\mbox{(NLL)}; 
 \alpha_s^2, \alpha_s v, v^2\,\mbox{(NNLL)}\bigg\}
 \,,
 \label{RNNLLorders}
\end{eqnarray}
where the indicated terms are of leading logarithmic (LL), next-to-leading
logarithmic (NLL), and next-to-next-to-leading logarithmic (NNLL) order.  The
summation of logarithms can be performed using renormalization group equations
in the framework of ``velocity NRQCD'' (vNRQCD)\
\cite{Luke1,amis,amis2,amis3}, an effective theory for heavy quark
pairs. The logarithmic corrections treated in this work are at NNLL order and
fully include the known NNLO QCD corrections from
Refs.~\cite{Hoang1,Melnikov1,Yakovlev1,Beneke1,Nagano1,Hoang2,Penin1}.  A
complete summation requires knowledge of the Wilson coefficients for potentials
and non-relativistic production currents at this order. For the NNLL potentials
the required Wilson coefficients have been calculated in
Refs.~\cite{amis,amis3,hms1}. The LL running of subleading $t\bar t$ production
currents is also required and is given below. For the leading production current
the Wilson coefficient is known at NLL~\cite{amis3} and
NNLO~\cite{Andrem1,Czarnecki1,Beneke4}. For this current only partial NNLL
results are known and an estimate is made for the uncertainty this induces in
our result.  The combination of known NNLL results significantly reduces the
size and scale dependence of previous NNLO QCD predictions~\cite{Hoang3}.

We emphasize that we do not treat electroweak effects consistently at the same
level as the QCD corrections, as implied by the counting $\Gamma_t\sim m_t v^2$
used in this paper. This would require the inclusion of non-factorizable
corrections as well as the single- and non-resonant background contributions,
which is beyond the scope of this work. Here we only include the top quark
width through an imaginary mass term, which can implemented by the
replacement\ \cite{Fadin1}, $E \, \equiv \, \sqrt{s}-2m_t \to \, E +
i\,\Gamma_t$ into a calculation for stable quarks.  For the total cross section
this represents a consistent next-to-leading order treatment of electroweak
effects~\cite{nonfactorizable}. We stress, however, that with the small
uncertainties in $R$ resulting from our NNLL QCD results, a systematic
treatment of higher order electroweak related effects is desirable.

The outline of the paper is as follows. In section~\ref{sectionresum} we review
an example from inclusive non-leptonic $B$-decays to emphasize some important
differences between fixed order and renormalization group improved
calculations. In section~\ref{sectionvnrqcd} we discuss the matching and running
of Wilson coefficients for top quark production near threshold.  In
section~\ref{sectioncrosssection} we evaluate the time ordered products of
currents at the endpoint of the renormalization group evolution. Pole mass
results are given in section~\ref{sectionpolemass}, while our implementation of
the $\SS$ mass is presented in section~\ref{sectionquarkmass}.  In
section~\ref{sectiondiscussion} the NNLL $\SS$ mass results are presented and
the remaining theoretical uncertainty is analyzed. The consequences of our
results for measurements of the top width, $\alpha_s(m_Z)$, and the Higgs top
Yukawa coupling are discussed in section~\ref{sectionphenomenology}, followed by
conclusions in section~\ref{sectionsummary}.

\section{Summing logarithms} \label{sectionresum}
 
To outline the steps in our renormalization group improved calculation it is
instructive to draw a parallel with the well-known computation for inclusive
non-leptonic $\bar B$-meson decays via the quark decay $b \to c \bar u d$. This
process involves several scales --- $m_W$, $m_b$, $m_c$ and $\lqcd$, but only
$m_W$ and $m_b$ will be discussed here.  The relevant weak decay Hamiltonian at
the scale $\mu$ is
\begin{eqnarray}
  H_W &=& {4G_F\over \sqrt{2}} V_{cb} V_{ud}^* \Bigg[C_1 \left(\mu\right)
  O_1 (\mu)\label{1.5.10}
 + C_2 \left( \mu \right) O_2 (\mu)
  \Bigg],\\[3pt]
  O_1 (\mu) &=& [\bilin {c^\alpha}{ \gamma_\mu P_L}{ b_\alpha}] [\bilin{ d^\beta}
  {\gamma^\mu P_L}{ u_\beta}],\qquad
  O_2 (\mu) = [\bilin {c^\beta}{ \gamma_\mu P_L}{ b_\alpha}] [\bilin{ d^\alpha}
  {\gamma^\mu P_L}{ u_\beta}] \,, \nn
\end{eqnarray}
where $P_L=(1-\gamma_5)/2$.  $H_W$ is $\mu$-independent and the $\mu$ dependence
of $C_i(\mu)$ is compensated by the $\mu$ dependence of $O_i(\mu)$. The
coefficients $C_{1,2}(m_W)$ are determined by integrating out $W$ bosons in the
$b \to c \bar u d$ amplitude at $\mu=m_W$
\begin{eqnarray}
 C_1 \left( \mu=m_W \right) &=& 1 + {\cal O} \left (\alpha_s \left(m_W \right)
 \right),\qquad
 C_2 \left( \mu=m_W \right) = 0 + {\cal O} \left (\alpha_s \left(m_W \right)
 \right).\label{1.5.15}
\end{eqnarray}
The coefficients can be computed using perturbation theory in $\alpha_s(m_W)$,
which is valid as long as $\alpha_s(m_W)$ is small. There are no large
logarithms in the matching calculation since we have chosen $\mu=m_W$.  The
decay amplitude can then be computed by taking the matrix elements of
$O_1\left(\mu=m_W \right)$ in the $\bar B$-meson. The problem is that these
matrix elements contain large logarithms of $m_W/m_b$. Using the renormalization
group these large logarithms can instead be moved into the coefficients $C_i$
and summed by scaling $O_i$ and $C_i$ from $\mu=m_W$ to $\mu=m_b$. Note that in
a fixed order expansion without the renormalization group large logarithms
exist even for $\mu=m_b$.  The renormalization group equation (RGE) for the
coefficients is
\begin{eqnarray} \label{adimc12}
 \mu{d\over d\mu} C_i = \gamma_{ji} C_j \,,\qquad\quad
 \gamma (g) = {g^2\over  8\pi^2} \left[\begin{array}{rr}
 -1 & 3\\ 3 & -1 \end{array} \right] + {\cal O}(g^4) \,, \label{1.5.24}
\end{eqnarray}
and is diagonalized by taking $C_\pm = C_1 \pm C_2$.  Provided the strong
coupling $\alpha_s (\mu)$ is small over the integration range in
Eq.~(\ref{adimc12}), higher order terms in $\gamma$ and $\beta$ can be neglected
and
\begin{eqnarray} \label{1.5.33}
 C_\pm \left(\mu\right) = \left[{\alpha_s
 (m_W)\over\alpha_s (\mu)}\right]^{a_{\pm}} \,,\qquad\quad
 a_+ = {2\over \beta_0} , \qquad a_- = -{4\over \beta_0}.
\end{eqnarray}
Using $\alpha_s(m_W)=0.12$ and $\alpha_s(m_b)=0.22$ the coefficients $C_\pm$ at
$\mu = m_b$ are
\begin{eqnarray}
 C_+(m_b) = 0.85, \qquad C_-(m_b)= 1.37 \,,
\end{eqnarray}
a change of -15\% and 37\% from their values at $m_W$.  Finally, the inclusive
$\bar B$ meson decay rate is computed from the imaginary part of the matrix
element of a time-ordered product, $\me{\bar B}{i\,\int d^4x\, \, T\:
H_W^\dagger(x)\ H_W(0)}{\bar B}]$.  At leading order in $1/m_b$ at the scale
$\mu=m_b$, the operator product expansion gives~\cite{vHQET}
\begin{eqnarray} \label{Gdc}
 \Gamma^{(\Delta c=1)} &=& {3G_F^2 m_b^5 \over 192 \pi^3}\abs{V_{cb}V_{ud}}^2 
 \left(C_1^2(m_b)+{2 \over 3} C_1(m_b) C_2(m_b) + C_2^2(m_b)\right)  
  f(m_c^2/m_b^2) \,,
\end{eqnarray}
where $f(m_c^2/m_b^2)$ is a phase-space factor. The QCD corrections to this
resummed result are of order $\alpha_s(m_b)$ with no large logarithm.

In the above example the computation of the decay rate was divided into three
parts:
\begin{enumerate}

\item Compute the coefficients $C_i\left( \mu=m_W \right)$ in a perturbation
series in $\alpha_s(m_W)$, which is valid as long as $\alpha_s(m_W)$ is small.

\item Scale $C_i\left( \mu \right)$ from $\mu=m_W$ to $\mu=m_b$ using the
renormalization group, which can be done provided $\alpha_s(\mu)$ is small in
the region $m_b \le \mu \le m_W$. This scaling sums terms of the form $[\alpha_s
\ln(m_b/m_W)]^k$. The terms neglected in the anomalous dimension are smaller
than those retained by $\alpha_s(\mu)$ at all points in the integration region.

\item Compute the decay rate using $H_W$ renormalized at $\mu=m_b$ in a
perturbation series in $\alpha_s(m_b)$ (and $1/m_b$), which is valid as long as
$\alpha_s(m_b)$ is small.

\end{enumerate}
Since $\alpha_s(\mu)$ is monotonically increasing as $\mu$ decreases, all three
steps above are valid as long as $\alpha_s(m_b)$ is small. However, it is
crucial to note that because of the renormalization group improvement we do not
need to assume that 
\begin{eqnarray}
  \alpha_s(m_W) \ln\bigg[ {m_W^2 \over m_b^2} \bigg] \simeq 0.7 
\end{eqnarray}
is a small expansion parameter. In contrast, a fixed order computation of the
decay rate in powers of $\alpha_s(\mu)$ contains logarithms of $m_W/\mu$ and
$m_b/\mu$, and is valid only if $\alpha_s(\mu)$ times these logarithms is
small. In particular, one needs $\alpha_s(m_b) \ln(m_W^2/m_b^2)$ to be small. As
a result, fixed order calculations have limited validity when processes involve
widely separated scales.

For non-relativistic top quarks the running spans an even larger range of
energies than the above example, $\mu\!=\! m_t$ down to $\mu\!=\! m_tv^2\sim
4\,{\rm GeV}$.  Therefore, at the order that anomalous dimensions depend
directly on the $m_t v^2$ energy scale (NNLO), it is not too surprising that
renormalization group improved perturbation theory is necessary to compute the
$\bar t t$ cross-section. Since $E\sim 4\,{\rm GeV}$, fixed order
perturbation theory would be valid only if
\begin{eqnarray} \label{ttbarlog}
 \alpha_s(m_t) \ln \bigg[ {m_t^2 \over E^2 } \bigg] \simeq 0.8
\end{eqnarray}
were small.  In contrast to Eq.~(\ref{ttbarlog}), renormalization group improved
perturbation theory only requires $\alpha_s(m_tv^2)$ to be small. Relative to
$B$-decays the radiative corrections for top quark production could be even
larger, since factors of $\alpha_s$ for the bound state do not always come with
a $1/\pi$. This is true of both the constant and logarithmically enhanced terms.

For the highest order computed, fixed order perturbation theory cannot
distinguish the scale at which to evaluate $\alpha_s$.  Mistaking an
$\alpha_s(m_t)$ for an $\alpha_s(m_t v^2)$ is a difference of a factor of two.
Renormalization group improved perturbation theory distinguishes the
$\alpha_s$'s that appear at each order. We note that in our NNLL order results,
the summation of logarithms which involve the $m_t v^2$ scale do tend to
numerically dominate.

\section{Effective Theory}
\label{sectionvnrqcd}

For our calculation we will employ vNRQCD\ \cite{Luke1,amis,amis2,amis3}, an
effective field theory which describes the non-relativistic dynamics of heavy
quark pairs, where the non-relativistic scales $m v$ (momentum) and $m v^2$
(energy) are larger than the hadronization scale $\Lambda_{\rm QCD}$, $m$ being
the heavy quark mass. This effective theory contains a consistent power counting
in $v$, so that a given Feynman diagram contributes to a single order in the $v$
expansion. It also accounts for the fact that the momentum (soft) and the energy
(ultrasoft) scales are correlated through the heavy quark equations of motions.
Finally, vNRQCD allows for the summation of logarithms of ratios of the hard
scale $m$ and the non-relativistic scales using a velocity renormalization
group. The most important momentum regions for heavy quark processes are those
with (energy, momentum) that are hard $(m,m)$, soft $(m v,m v)$, potential $(m
v^2,m v)$, and ultrasoft $(m v^2,m v^2)$ \ \cite{Beneke2}.  The effective theory
is formulated by including only those quark and gluonic degrees of freedom which
can become on-shell for scales below $m$. The on-shell degrees of freedom are
gluons and massless quarks with soft and ultrasoft energies and momenta, and
heavy quarks with potential energies and momenta. All off-shell effects such as
those from hard quarks and gluons, potential gluons, and soft quarks are
accounted for by on-shell matching of vNRQCD to full QCD at the hard scale.

\begin{figure}[t] 
\begin{center}
 \leavevmode
 \epsfxsize=12cm
 \leavevmode
 \epsffile[65 535 565 605]{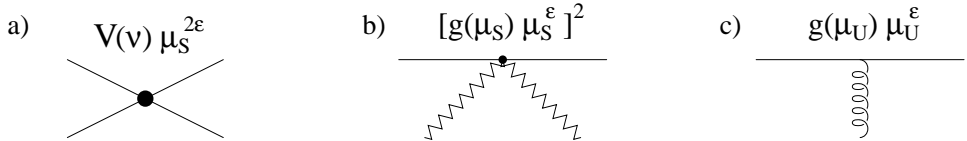}
 \vskip  0.5cm
 \caption{Examples of potential (a), soft (b), and ultrasoft (c) interactions 
 in vNRQCD.
 \label{fig:fr} }
\end{center}
\end{figure}

Only ultrasoft energies and momenta are treated as continuous variables, and
ultrasoft momenta are separated from larger momenta by employing the multipole
expansion. Soft energies and momenta appear as discrete indices for potential
quarks and soft gluons and must be summed over, similar to the velocity index
$v^\mu$ for the static quark in HQET\ \cite{vHQET}. For instance, the first few
terms that are bilinear in the quark field $\psip p$ are
\begin{eqnarray} \label{Lke}
 {\mathcal L}(x) &=& \sum_{\sitbf{p}}
   \psip p ^\dagger(x)   \Biggl\{ i \partial^0 - {{\itbf{p}}^2 \over 2 m} 
   +\frac{{\itbf{p}}^4}{8m^3} \Biggr\} \psip p(x), 
\end{eqnarray}
where $m$ is the pole mass, and there is a similar equation for the anti-quark
field $\chip p$.  Sums over soft and potential energies and momenta in
combination with ultrasoft integrations in loop integrals can be rewritten as
continuous loop integrals of soft and potential energies and momenta.  The
vNRQCD Lagrangian also contains operators that describe potential-type 4-quark
interactions (Fig.~\ref{fig:fr}a) originating from potential gluons and other
off-shell modes, interactions of quarks with soft gluons (Fig.~\ref{fig:fr}b),
and ultrasoft gluons (Fig.~\ref{fig:fr}c), as well as self-interactions of soft
and ultrasoft gluons. The potential and soft interactions depend non-locally on
the soft indices, but are local in ultrasoft momenta.  Gauge invariance in soft
energies and momenta is recovered through reparameterization invariance. Loop
integrals are regularized with dimensional regularization in $d=4-2\epsilon$
dimensions using the $\overline{\mbox{MS}}$ subtraction scheme.
%
%

The crucial feature of vNRQCD is that the dimensionful $\overline{\mbox{MS}}$
subtraction scale for loop integrations over soft and potential momenta is
$\mu_S$, whereas for loop integrations with ultrasoft momenta it is $\mu_U$.
Within dimensional regularization the two different subtraction scales are
necessary to properly recover the full $d$-dimensional non-relativistic phase
space~\cite{Luke1}. Powers of $(\mu_U)^{\epsilon}$ and $(\mu_S)^{\epsilon}$ are
determined uniquely by the interaction vertices~\cite{amis2} as shown in
Fig.~\ref{fig:fr}. Furthermore, the two scales are correlated, $\mu_S=m \nu$ and
$\mu_U=m \nu^2$, where the dimensionless velocity subtraction point $\nu$ takes
the role of the $\overline{\mbox{MS}}$ scaling parameter.  The correlation
between $\mu_S$ and $\mu_U$ is an inherent property of vNRQCD, since the energy
and momentum scales are related through the quark equations of motion.  For
NRQED this correlation is necessary to reproduce through running the
$(\ln\alpha)^k$, $k\ge 2$ contributions in Lamb shifts, hyperfine splittings and
the corrections to the ortho- and para-positronium decay
widths~\cite{amis4,Manohar6}.  The renormalization group equations for Wilson
coefficients are determined in the canonical way with the exception that they
are formulated in terms of the subtraction velocity $\nu$.

Lowering $\nu$ to a scale $v_0$ of order the quark velocity sums all logarithms
of the soft and ultrasoft scales into the Wilson coefficients of the vNRQCD
operators. At NNLL order this includes logarithms originating from ultrasoft
radiation effects.  The $t\bar t$ state is predominantly Coulombic, so
$v_0\simeq C_F \alpha_s\simeq 0.15$--$0.20$. Once $\nu$ is lowered to $v_0$,
power counting shows that matrix elements with ultrasoft gluons no longer have
to be taken into account at NNLL for the description of a heavy quark pair.
Thus, at $\nu= v_0$ the NNLL Green functions of a color singlet quark-antiquark
system are described by the two-body Schr\"odinger equation with relativistic
corrections.

In the remainder of this section we introduce the Wilson coefficients which are
matched at the hard scale ($\nu=1$), and run from $\nu=1$ to $v_0$.

\subsection{Potentials}

The potentials in the NNLL Schr\"odinger equation arise from the four-quark
matrix elements of the potential-type operators~\cite{Pineda} and from
time-ordered products of operators describing interactions with soft gluons. The
latter account, for example, for the radiative corrections to the Coulomb
potential, and include the dependence of the Schr\"odinger potentials on
logarithms of soft momenta.  Adding these contributions, the effective order
$1/v$ Coulomb potential is
\begin{eqnarray}
 \tilde V_c(\bmp,\bmq) 
 & = &
 \frac{{\cal{V}}_c^{(s)}(\nu)}{\bmk^2}\, 
 -\,\frac{4\pi C_F\, \alpha_s(m\nu)}{\bmk^2}\, \Bigg\{\,
 \frac{\alpha_s(m\nu)}{4\pi}\,\bigg[\,
 -\beta_0\,\ln\Big(\frac{\bmk^2}{m^2\nu^2}\Big) + a_1
 \,\bigg] \nonumber\\[2mm] 
& & \quad + \bigg(\frac{\alpha_s(\mu_s)}{4\pi}\bigg)^2\,\bigg[\,
 \beta_0^2\,\ln^2\Big(\frac{\bmk^2}{m^2\nu^2}\Big)  
 - \Big(2\,\beta_0\,a_1 +
 \beta_1\Big)\,\ln\Big(\frac{\bmk^2}{m^2\nu^2}\Big) + a_2 \,\bigg]
 \,\Bigg\} \,,
 \label{VCoulomb}
\end{eqnarray}
where $\bmk=\bmp-\bmq$ is the momentum transfer.  The first term in
Eq.~(\ref{VCoulomb}) has a coefficient ${\cal V}_c^{(s)}(\nu)$ that contains the
summation of NNLL logarithms coming from soft and ultrasoft gluons and was
determined in Ref.\ \cite{hms1}. The second term contains the one- and two- loop
corrections to the Coulomb potential\ \cite{Schroder1}. In vNRQCD they arise
from time-ordered products of the lowest order operators describing the
interaction of quarks with soft gluons~\cite{hms1}. The couplings in these
interactions are simply $\alpha_s(\mu_S)=\alpha_s(m\nu)$ and evolve with the QCD
$\beta$-function.

At NNLL order a virtual photon or $Z$ boson produces top quarks in $^3S_1$ and
$^3P_1$ states through the vector and axial-vector currents respectively. We can
simplify the spin dependence of the subleading $1/m^2$ potentials by projecting
onto the spin triplet channel. To do this we choose the scheme where traces over
$\sigma^i$ matrices are done in three dimensions (using ${\bf S}^2=2$ for
triplet states).\footnote{The difference between using three and $d$ dimensional
$\sigma$ matrices is simply a change in renormalization scheme. A similar scheme
dependence arises in chiral perturbation theory~\cite{Savage}.} The order
$v^{0}$ and $v^1$ potentials are\,\footnote{At NNLO the potentials $\tilde
V_\delta$, $\tilde V_r$ and $\tilde V_k$ are equivalent to what was referred to
as ``Breit-Fermi'' and ``non-Abelian'' potentials in some publications on the
NNLO corrections (see e.g.\ \cite{Hoang1,Melnikov1,Yakovlev1}).}
\begin{eqnarray}  \label{Vkdr}
 \tilde V_k(\bmp,\bmq) & = & \frac{\pi^2}{m |\bmk|}\, 
    {\cal{V}}_k^{(s)}(\nu) \,, \\[2mm]
 \tilde V_\delta(\bmp,\bmq) & = & 
\frac{{\cal{V}}_2^{(s)}(\nu) 
    +2 {\cal{V}}_s^{(s)}(\nu)}{m^2} \,, \nn \\[2mm]
 \tilde V_r(\bmp,\bmq) & = & \frac{(\bmp^2+\bmq^2)}{2 m^2 \bmk^2}\, 
    {\cal{V}}_r^{(s)}(\nu) \,. \nn
\end{eqnarray}

In vNRQCD the potential coefficients at $\nu=1$ are obtained with on-shell
matching at the hard scale. The potentials $\tilde V_\delta$ and $\tilde V_r$
are of order $\alpha_s$ and lead to terms in the cross section that are
$v^2$-suppressed. Their evolution therefore only needs to be taken into account
in LL order and was determined in Ref.\ \cite{amis}. The potential $\tilde V_k$
arises only at order $\alpha_s^2$ and also leads to terms in the cross section
that are $v^2$-suppressed. Its evolution needs to be known at NLL and was
determined in Ref.\ \cite{amis3}.  At NNLL order the complete sum of
potential-like interactions is
\begin{eqnarray}
 \tilde V(\bmp,\bmq) \, = \, 
 \tilde V_c(\bmp,\bmq) + 
 \tilde V_\delta(\bmp,\bmq) + 
 \tilde V_r(\bmp,\bmq) +
 \tilde V_k(\bmp,\bmq)  
 \,.
\label{Vsdetail}
\end{eqnarray} 
Explicit formul\ae\ for the Wilson coefficients and constants in Eqs.\
(\ref{VCoulomb}) and (\ref{Vkdr}) are collected in Appendix A.
 
\subsection{Currents} \label{eftc}

To describe quark-antiquark production close to threshold at NNLL order we need
to know the Wilson coefficient at NNLL order of the dimension three ${}^3S_1$
production current, and at LL order the dimension four ${}^3P_1$ and dimension
five ${}^3S_1$ currents. The vector production current is ${\bf
J}^v_{\sitbf{p}}= c_1 \O{p}{1} + c_2 \O{p}{2}$, where
\begin{eqnarray}\label{Ov}
 \O{p}{1} & = & {\psip{p}}^\dagger\, \bsigma(i\sigma_2)\, {\chip{-p}^*} \,, 
   \\[2mm]
 \O{p}{2} & = & \frac{1}{m^2}\, {\psip{p}}^\dagger\, 
    \bmp^2\bsigma (i\sigma_2)\, {\chip{-p}^*} \,, \nn
\end{eqnarray} 
and the relevant axial-vector current is ${\bf J}^a_{\sitbf{p}}=
c_3 \O{p}{3} $, where
\begin{eqnarray}\label{Oa}
 \O{p}{3} & = & \frac{-i}{2m}\, {\psip{p}}^\dagger\, 
      [\,\bsigma,\bsigma\cdot\bmp\,]\,(i\sigma_2)\,
   {\chip{-p}^*} \,. 
\end{eqnarray} 
In Eqs.~(\ref{Ov}) and (\ref{Oa}) $\mbox{\boldmath $p$}$ is the soft-momentum
index of the quarks. The corresponding annihilation currents
${\O{p}{1-3}}^\dagger$ are obtained by complex conjugation. In the basis of
operators we are using there is an additional dimension five vector current,
\begin{eqnarray} \label{OvD}
 \O{p}{4} & = & \frac{1}{m^2}\, {\psip{p}}^\dagger\, 
   \bigg(\, \bmp (\bsigma\cdot\bmp)-\bsigma\,\frac{\bmp^2}{3}\, \bigg)
   \,(i\sigma_2)\,{\chip{-p}^*} \,.
\end{eqnarray}
However it produces a $D$-wave quark-antiquark pair and therefore does not
contribute at NNLL order.  

The matching condition at the hard scale for the Wilson coefficient $c_1(\nu=1)$
is needed to order $\alpha_s^2$.  Its value is determined from matching the
two-loop result for the quark-antiquark production amplitude close to threshold
in full QCD to the corresponding amplitude in vNRQCD and is given in Appendix A.
Since the two-loop result for $c_1(1)$ is scheme dependent, our result is
different from that obtained with the threshold expansion\ \cite{Beneke2}. The
LL anomalous dimension for $c_1$ is zero.  The evolution of $c_1$ for $\nu<1$ at
NLL order was determined analytically in Ref.\ \cite{amis3}. Due to the length
of the resulting expression this analytic result for $\nu<1$ is not repeated
here.  At NNLL order the evolution of $c_1(\nu)$ requires the three loop
anomalous dimension which is not completely known. In Appendix A we discuss the
form of contributions to this anomalous dimension, and investigate their effect
on $c_1(\nu)$. Based on the contributions that are currently known, and a
parametric estimate for the size of the unknown contributions, we find that the
uncertainty in $c_1(\nu)$ from neglecting this running is likely to be at a
level $\lesssim 1\%$.

%
\begin{figure}[t!] 
\begin{center}
 \leavevmode
 \epsfxsize=3.cm
 \leavevmode
 \epsffile[63 412 213 502]{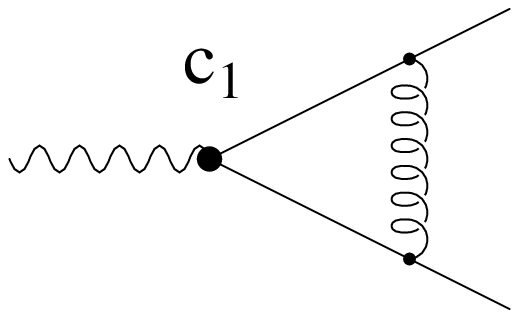}
 \hskip 1cm
 \epsfxsize=3.cm
 \epsffile[63 412 213 502]{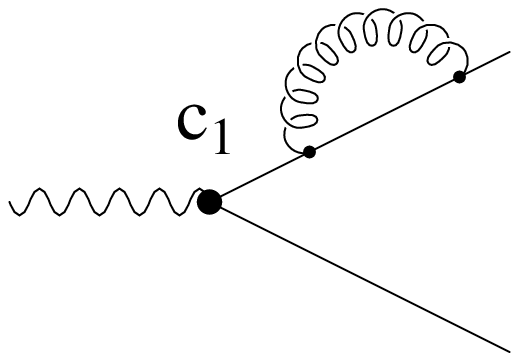}
 \hskip 1cm
 \epsfxsize=3.cm
 \epsffile[63 412 213 502]{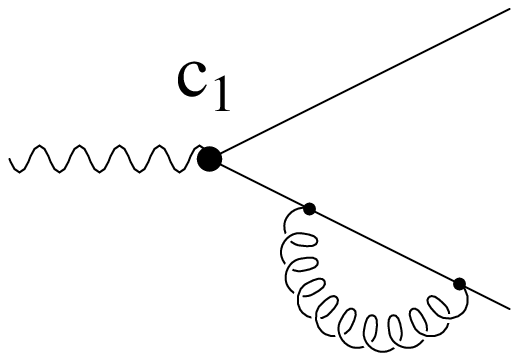} 
 \vskip  0.2cm
 \caption{Graphs with an ultrasoft gluon with ${\bmp}\cdot {\bf A}$ couplings 
 which contribute to the running of $c_2(\nu)$.
 \label{fig:c2} }
\end{center}
\end{figure}
The coefficients $c_2$ and $c_3$ are needed at LL order. This involves tree
level matching, together with one-loop running.  At tree level expanding the
full theory currents, $\bar t \bgamma t$ and $\bar t \bgamma\gamma_5 t$ gives
\begin{eqnarray} \label{bcc23}
  c_2(\nu=1) = -1/6 \,, \qquad\qquad c_3(\nu=1)=1 \,.
\end{eqnarray}
The LL evolution of $c_3$ would be determined by divergent order $\alpha_s v$
graphs, but no such diagrams exist so $c_3(\nu)=1$.  For $c_2(\nu)$ we need
the divergent order $\alpha_s v^2$ graphs which are shown in Fig.~\ref{fig:c2}.
Together they give the anomalous dimension
\begin{eqnarray} \label{adc2}
 \nu{\partial\over\partial\nu} c_2(\nu) &=& \frac{8C_F}{3\pi} 
   \alpha_s(m\nu^2)\, c_1(\nu) \,.
\end{eqnarray}
Integrating Eq.~(\ref{adc2}) with the boundary condition in Eq.~(\ref{bcc23})
and the LL value of $c_1(\nu)$ gives the LL solution for $c_2(\nu)$:\footnote{
Eq.~(\ref{c2ll}) was used in Ref.~\cite{Hoang8}, but corrects a typo in Eq.~(13)
of this reference.}
\begin{eqnarray} \label{c2ll}
 c_2(\nu)&=& -\frac{1}{6}-
 \frac{8 C_F}{3\beta_0} \ln\bigg[ \frac{\alpha_s(m\nu^2)}
 {\alpha_s(m)} \bigg] \,.
\end{eqnarray}
A check of this result is obtained by expanding in $\alpha_s(m)$ 
\begin{eqnarray}
 c_2(\nu)&=& -\frac{1}{6}+
 \frac{4 C_F}{3\pi}\,\alpha_s(m)\,\ln\big(\frac{m\nu^2}{m} \big) + \ldots \,,
\end{eqnarray}
where the $\ln(\nu^2)$ term agrees with the $\ln(\mu/m)$ term derived in
Ref.~\cite{Luke2}. (Note that the operator $\O{p}{2}$ in Ref.~\cite{Luke2} was
defined with the opposite sign.)

\subsection{Width and Nonperturbative Effects}

To incorporate the effect of the large top quark width we include in the 
effective Lagrangian the operators
\begin{eqnarray} \label{Gtop}
 {\cal L} = \sum_{\sitbf{p}} \psip{p}^\dagger \: \frac{i}{2} \Gamma_t\: 
  \psip{p} \ +\  
  \sum_{\sitbf{p}}   \chip{p}^\dagger\: \frac{i}{2} \Gamma_t\: \chip{p} \,.
\end{eqnarray}
In the Standard Model the dominant decay channel is $t\to bW^+$ and gives a
width of $\Gamma_t=1.43\,{\rm GeV}$ (including one-loop
electroweak~\cite{Beenakker1} and two-loop QCD~\cite{Czarnecki2} effects). For
power counting we will treat $\Gamma_t$ as a fixed number to be determined from
experiment. The typical energy of a Coulombic top quark is $E\sim mv^2\sim
4\,{\rm GeV}$, and we take $\Gamma_t\sim E$. Thus, the propagator for a single
top (or antitop) with momentum $(p^0,{\itbf{p}})$ is
\begin{eqnarray}
  \frac{i}{p^0 - {\itbf{p}}^2/(2m) + i\Gamma_t/2 + i\epsilon} \,.
\end{eqnarray}
In our results the use of this propagator is equivalent to the replacement $E\to
E+i\Gamma_t$ in $t\bar t$ Green functions from the stable quark calculation.
For the total $t \bar t$ cross section in the Coulombic regime the operator in
Eq.~(\ref{Gtop}) is sufficient to account for all electroweak effects at
LO~\cite{Fadin1} and NLO~\cite{nonfactorizable}. The complete set of electroweak
effects at NNLL order is currently unknown.

The large top quark width plays a crucial role in suppressing the size of
nonperturbative hadronic contributions governed by the scale $\Lambda_{\rm
QCD}$. Even at threshold ($E=0$), the $t\bar t$ system has an effective energy
set by the perturbative scale $\Gamma_t$. For stable quarks, an operator product
expansion~\cite{Voloshin1,Leutwyler1} in powers of $\Lambda_{\rm QCD}/E$ can be
used to incorporate non-perturbative contributions.  Naively, the size of the
first non-perturbative correction for $t\bar t$ production is therefore
$[\Lambda_{\rm QCD}/(E+i\Gamma_t)]^4$ from the gluon condensate operator.
However, because the momenta of the gluons, of order $\Lambda_{\rm QCD}$, is
much smaller than the momenta of the quarks $mv\sim 25\,{\rm GeV}$, their
coupling to quarks is governed by a multipole expansion. The lowest order
interaction is $[\psip{p}^\dagger \,{\itbf{p}}\cdot {\bf A} \psip{p}]$,
giving an additional $v^2$ suppression~\cite{Beneke5}.  The size of the leading
nonperturbative effect is therefore $\delta R^{np}/R\sim \Lambda_{\rm QCD}^4 v^2
/(E+i\Gamma_t)^4\lesssim 10^{-4}$.  This is consistent with the size of
the gluon condensate corrections to $R$ computed in Ref.~\cite{Fadin2}, and is
more than two orders of magnitude smaller than the corrections considered here.

\section{Total Cross Section}
\label{sectioncrosssection}

In this section we compute the time ordered products of currents at a scale
corresponding to the typical momentum and energy of the non-relativistic quarks,
namely $\nu\simeq 0.15-0.20$.  In full QCD the expression for the total cross
section $\sigma_{\rm tot}^{\gamma,Z}(e^+e^-\to \gamma^*,Z^*\to q\bar q)$ for
quarks at center of mass energy $\sqrt{s}$ is
\begin{eqnarray}
  \sigma_{\rm tot}^{\gamma,Z}(s) = \sigma_{\rm pt} 
  \Big[\, F^v(s)\,R^v(s) +  F^a(s) R^a(s) \Big] \,,
\label{totalcross}
\end{eqnarray}
where $\sigma_{\rm pt}=4\pi\alpha^2/(3 s)$. The vector and axial-vector
$R$-ratios are
\begin{eqnarray} \label{fullR}
 R^v(s) \, =  \,\frac{4 \pi }{s}\,\mbox{Im}\,\left[-i\int d^4x\: e^{i q\cdot x}
  \left\langle\,0\,\left|\, T\, j^v_{\mu}(x) \,
  {j^v}^{\mu} (0)\, \right|\,0\,\right\rangle\,\right] \,, \nn\\[2pt]
 R^a(s) \, =  \,\frac{4 \pi }{s}\,\mbox{Im}\,\left[-i\int d^4x\: e^{i q\cdot x}
  \left\langle\,0\,\left|\, T\, j^a_{\mu}(x) \,
  {j^a}^{\mu} (0)\, \right|\,0\,\right\rangle\,\right] \,, 
\end{eqnarray}
where $q=(\sqrt{s},0)$ and $j^{v}_\mu$ ($j^{a}_\mu$) is the vector
(axial-vector) current that produces a quark-antiquark pair. With both $\gamma$
and $Z$ exchange the prefactors in Eq.~(\ref{totalcross}) are
\begin{eqnarray}
  F^v(s) &=& \bigg[\, Q_q^2 - 
   \frac{2 s\, v_e v_q Q_q}{s-m_Z^2} + 
   \frac{s^2 (v_e^2+a_e^2)v_q^2}{(s-m_Z^2)^2}\, \bigg]\,,
\qquad 
  F^a(s) = \frac{s^2\, (v_e^2+a_e^2)a_q^2}{ (s-m_Z^2)^2 } \,,
\end{eqnarray}
where 
\begin{eqnarray}
  v_f = \frac{T_3^f-2 Q_f \sin^2\theta_W}{2\sin\theta_W \cos\theta_W}\,,
  \qquad\qquad
  a_f = \frac{T_3^f}{2\sin\theta_W \cos\theta_W} \,.
\end{eqnarray}
Here $Q_f$ is the charge for fermion $f$, $T_3^f$ is the third component of weak
isospin, and $\theta_W$ is the weak mixing angle.

In vNRQCD at NNLL order the current correlators are replaced by the correlators
of the non-relativistic currents $\O{p}{i}$, so that
\begin{eqnarray} \label{Rveft}
 R^v(s) & = & \frac{4\pi}{s}\,
 \mbox{Im}\Big[\,
 c_1^2(\nu)\,{\cal A}_1(v,m,\nu) + 
 2\,c_1(\nu)\,c_2(\nu)\,{\cal A}_2(v,m,\nu) \,\Big] \,,
\\[4mm] \label{Raeft}
 R^a(s) & = &  \frac{4\pi}{s}\,
 \mbox{Im}\Big[\,c_3^2(\nu)\,{\cal A}_3(v,m,\nu)\,\Big] \,,
\end{eqnarray}
with  
\begin{eqnarray}
 {\cal A}_1(v,m,\nu) & = & i\,
 \sum\limits_{\mbox{\scriptsize\boldmath $p$},\mbox{\scriptsize\boldmath $p'$}}
 \int\! d^4x\: e^{i \hat{q} \cdot x}\:
 \Big\langle\,0\,\Big|\, T\, \O{p}{1}(x){\Od{p'}{1}}(0)
 \Big|\,0\,\Big\rangle \,, \nn
\\[2mm]
 {\cal A}_2(v,m,\nu) & = &
 \frac{i}{2}\, 
 \sum\limits_{\mbox{\scriptsize\boldmath $p$},\mbox{\scriptsize\boldmath $p'$}}
 \int\! d^4x\: e^{i \hat{q}\cdot x}\:
 \Big\langle\,0\,\Big|\,
 T\,\Big[ \O{p}{1}(x){\Od{p'}{2}}(0)+\O{p}{2}(x){\Od{p'}{1}}(0)
 \Big] \Big|\,0\,\Big\rangle \,, \nn
\\[2mm]
 {\cal A}_3(v,m,\nu) & = & i\, 
 \sum\limits_{\mbox{\scriptsize\boldmath $p$},\mbox{\scriptsize\boldmath $p'$}}
 \int\! d^4x\: e^{i \hat{q}\cdot x}\:
 \Big\langle\,0\,\Big|\, T\, \O{p}{3}(x){\Od{p'}{3}}(0)\Big|\,0\,\Big\rangle \,.
\end{eqnarray}
Here $\hat{q}\equiv(\sqrt{s}-2m,0)$, and the current operators $\O{p}{i}$ are
defined in Eqs.~(\ref{Ov}) and (\ref{Oa}).  The correlators ${\cal A}_i$ are
functions of the quark mass $m$, velocity renormalization scale $\nu$, and the
effective velocity
\begin{eqnarray} \label{vdefwidth}
  v & = &
 \left(\frac{\sqrt{s}-2m+i\Gamma_t}{m}\right)^{\frac{1}{2}} \,.
\end{eqnarray}
Here $m$ is the pole mass.  For our calculation evaluating matrix elements at
the low-scale corresponds to computing the ${\cal A}_i$'s with $\nu$ of the
order a typical Coulombic top quark's velocity.  The correlator ${\cal A}_2$ can
be related to ${\cal A}_1$ using the quark equation of motion giving
\begin{eqnarray}
  {\cal A}_2(v,m,\nu) & = & {v^2}\,{\cal A}_1(v,m,\nu) \,.
\end{eqnarray}
Performing the spin traces in 3-dimensions the correlators ${\cal A}_1$ and
${\cal A}_3$ are determined by the zero-distance coordinate space Green
functions,
\begin{eqnarray}
{\cal A}_1(v,m,\nu) 
 \, &=& \, 6\,N_c\, \mu_S^{4\epsilon} \int D^n\bmp\,D^n\bmp^\prime\, 
      \tilde G_{v,m,\nu}(\bmp,\bmp^\prime)  \,, \nn\\[2mm]
{\cal A}_3(v,m,\nu) 
 \, &=& \, \frac{12\,N_c}{m^2\:n}\,\mu_S^{4\epsilon} \int D^n\bmp\,
      D^n\bmp^\prime\, (\bmp\cdot \bmp^\prime)\,
      \tilde G_{v,m,\nu}(\bmp,\bmp^\prime)  \,.
\label{Acorrelators}
\end{eqnarray}
Here $n\equiv d\!-\!1=3\!-\!2\epsilon$, we use $D^n\mbox{\boldmath $k$}\equiv
e^{\epsilon\gamma_E} (4\pi)^{-\epsilon}\,d^n\bmk/(2\pi)^n$ which converts
$\mu_S$ (and $\nu$) from the MS to the $\overline{\rm MS}$ scheme, and $N_c=3$
is the number of colors.  The dependence of the ${\cal A}_3$ prefactor on $n$
comes from taking the trace over $\bsigma$ matrices in the $P$-wave current in
$3$ dimensions, and then projecting the dot product ${\bmp}\cdot {\bmp}\,'$ back
into $n$-dimensions.

To determine the correlators ${\cal A}_{1}$ and ${\cal A}_{3}$ at NNLL order we
use a combination of numerical and analytic calculations. Ultraviolet divergent
contributions are computed with dimensional regularization in the $\overline{\rm
MS}$ scheme, which is required for consistency with the scheme used to compute
the matching and running of the Wilson coefficients. Our strategy for the
analytic calculations for these potentials follows that in Ref.\
\cite{Hoang1}. In momentum space the NNLL Schr\"odinger equation for the Green
function for a color singlet quark-antiquark pair is
\begin{eqnarray} 
 && 
 \bigg[ \frac{\bmp^2}{m} - \frac{\bmp^4}{4m^3} - \tilde E \bigg]
 \, \tilde G_{s,m,\nu}(\bmp,\bmp^\prime)
 + \int D^n\bmk\, \mu_S^{2\epsilon}\, \tilde V(\bmp,\bmk)\, 
 \tilde G_{s,m,\nu}(\bmk,\bmp^\prime)
 = \, (2\pi)^n\,\delta^{(n)}(\bmp-\bmp^\prime) \,,
\label{NNLLSchroedinger}
\end{eqnarray}
where $\tilde E\equiv mv^2$ and $\tilde V({\itbf{p}},{\itbf{k}})$ is given in
Eq.~(\ref{Vsdetail}).

We start with the vector correlator ${\cal A}_{1}$.  The lowest order
$S$-wave Coulomb Green function at zero distances reads
\begin{eqnarray}
 G^0(a,v,m,\nu) & = &
 \frac{m^2}{4\pi}\left\{\,
 i\,v - a\left[\,\ln\left(\frac{-i\,v}{\nu}\right)
 -\frac{1}{2}+\ln 2+\gamma_E+\psi\left(1\!-\!\frac{i\,a}{2\,v}\right)\,\right]
 \,\right\}
 \nonumber \\ & &
 +\,\frac{m^2\,a}{4 \pi}\,\,\frac{1}{4\,\epsilon}
\label{deltaGCoul}
\end{eqnarray}  
where $a\equiv -{\cal V}_c(\nu)/(4\pi)$. Eq.~(\ref{deltaGCoul}) is a standard
result~\cite{Coulomb}, except for the scheme dependent constants entering with
the divergence. This term was computed from the two-loop graph with a single
${\cal V}_c$ insertion, and Eq.~(\ref{deltaGCoul}) agrees with the \ms result
obtained in Ref.\ \cite{Soto1}.  The $1/\epsilon$ pole is canceled by a
counterterm for the time ordered product of currents. An identical divergence
structure appears with the full $\tilde V_c$ potential in Eq.~(\ref{VCoulomb}).
Since this divergence and corresponding constants appear only in the real part
of $G^0$, for the purpose of computing ${\rm Im}[{\cal A}_1]$ it does not matter
whether or not they are subtracted in $\overline{\rm MS}$. Therefore, we
determine the contribution from the Coulomb potential by the exact solution of
the equation
\begin{eqnarray} 
\bigg[
 \frac{\bmp^2}{m} - E \bigg]\,
 \tilde G^c_{v,m,\nu}(\bmp,\bmp^\prime)
 + \int \frac{d^3\bmk}{(2\pi)^3}\,
 \tilde V_c(\bmp,\bmk)\,
 \tilde G^c_{v,m,\nu}(\bmk,\bmp^\prime)
 \, = \, 
 (2\pi)^3\,\delta^{(3)}(\bmp-\bmp^\prime)  \,,
\label{CoulombSchroedinger}
\end{eqnarray}
using numerical techniques developed in Refs.\ \cite{Strassler1,Jezabek1}. We
will denote by $G^c(a,v,m,\nu)$ the solution for the zero-distance coordinate
space Green function obtained from $\tilde G^c(v,m,\nu)(\bmp,\bmp^\prime)$.  

On the other hand, the first order perturbative corrections to ${\cal
A}_1(v,m,\nu)$ from the potentials $\tilde V_k$, $\tilde V_\delta$, and $\tilde
V_r$, and the kinetic energy correction, ${\itbf{p}}^4/(4 m^3)$, have
ultraviolet sub-divergences which must be subtracted in $\overline{\rm MS}$.
For the first order corrections to the zero-distance Green function in
dimensional regularization from the kinetic energy correction and the potentials
$\tilde V_k$, $\tilde V_\delta$, $\tilde V_r$ (leaving out the Wilson
coefficients) we find:
\begin{eqnarray} \label{deltaGkinetic}
 \delta G^{\rm kin}(a,v,m,\nu) 
 & = & 
 \frac{a\,m^2}{16\pi}\left\{\,
 i\,v - a\left[\,\ln\left(\frac{-i\,v}{\nu}\right)
 -\frac{3}{2}+\ln 2+\gamma_E+\psi\left(1-\frac{i\,a}{2\,v}\right)\,\right]
 \,\right\}^2
 \\* & &
 -\,\frac{m^2}{16 \pi}\,\left[\, -a\,v^2 + 
 \frac{a^3}{4}\left(\frac{1}{4\,\epsilon^2}-\frac{1}{\epsilon}-2
 \right)
 \,\right]
 \nonumber \\* & &
 +\frac{v^2}{2}\left(1
 +a\,\frac{\partial}{\partial a}
 +\frac{v}{4}\,\frac{\partial}{\partial v}
 \right)\,G^0(a,v,m,\nu) \,,
 \nn \\[2mm]
 \delta G^k(a,v,m,\nu) 
 & = & 
 -\frac{m^2}{8 \pi\, a}\left\{\,
 i\,v - a\left[\,\ln\left(\frac{-i\,v}{\nu}\right)
 -2+2\ln 2+\gamma_E+\psi\left(1-\frac{i\,a}{2\,v}\right)\,\right]
 \,\right\}^2
  \\ 
 &+& \,\frac{m^2}{8 \pi\,a}\,\left[\, -v^2 + 
 \frac{a^2}{4}\,\left(\frac{1}{3\,\epsilon^2}
 -\frac{2}{\epsilon}\left(1-\frac{2}{3}\ln 2\right)+\frac{4}{3}
 - 8\ln 2+\frac{8}{3}\ln^2 2+\frac{\pi^2}{9}\right)
 \,\right], \label{deltaGk}
 \nonumber \\[2mm] 
 \delta G^\delta(a,v,m,\nu) 
 & = & 
 -\frac{m^2}{16\pi^2}\left\{\,
 i\,v - a\left[\,\ln\left(\frac{-i\,v}{\nu}\right)
 -\frac{1}{2}+\ln 2+\gamma_E+\psi\left(1-\frac{i\,a}{2\,v}\right)\,\right]
 \,\right\}^2
 \nonumber \\* & &
 +\,\frac{m^2\,a^2}{256 \pi^2}\,\,\frac{1}{\epsilon} \,,
 \label{deltaGdelta}
\\[2mm]
 \delta G^r(a,v,m,\nu) 
 & = & 
 -\frac{m^2}{16\pi^2}\left\{\,
 i\,v - a\left[\,\ln\left(\frac{-i\,v}{\nu}\right)
 -\frac{3}{2}+\ln 2+\gamma_E+\psi\left(1-\frac{i\,a}{2\,v}\right)\,\right]
 \,\right\}^2
 \nonumber \\* & &
 +\,\frac{m^2}{16 \pi^2}\,\left[\, -v^2 + 
 \frac{a^2}{4}\left(\frac{1}{4\,\epsilon^2}-\frac{1}{\epsilon}-2
\right)
 \,\right]
 - \frac{v^2}{4\pi}\,\frac{\partial}{\partial a}\,G^0(a,v,m,\nu) \,.
\label{deltaGr}
\end{eqnarray}
In deriving these equations we have included the counterterms generated by
renormalizing the $\O{p}{1}$ current~\cite{Luke1} at NLL order.  These
counterterm graphs are sufficient to cancel all subdivergences.  The remaining
overall divergences in Eqs.~(\ref{deltaGkinetic}) to (\ref{deltaGr}) are of the
form $1/\epsilon$ and $v^2/\epsilon$ and are canceled by vacuum counterterms for
the current correlator, and  can be dropped.  The final renormalized
expression for the NNLL vector correlator ${\cal A}_1$ is then
\begin{eqnarray}
{\cal A}_1(v,m,\nu)
 & = &
 6 \,N_c\,\bigg[\,
 G^c(a,v,m,\nu)
 + \left({\cal{V}}_2^{(s)}(\nu)+2{\cal{V}}_s^{(s)}(\nu)\right)\, 
 \delta G^\delta(a,v,m,\nu) \nonumber
\\[2mm] & & \hspace{0.1cm}
+ \,{\cal{V}}_r^{(s)}(\nu)\,\delta G^r(a,v,m,\nu) 
+ \,{\cal{V}}_k^{(s)}(\nu)\, \delta G^k(a,v,m,\nu) 
+ \,\delta G^{\rm kin}(a,v,m,\nu) 
\,\bigg]
\,.
\label{NNLLcrosssection}
\end{eqnarray}

For the computation of the axial-vector correlator ${\cal A}_3$ at NNLL we only
need to consider the leading order Green function because the axial-vector
current is already suppressed by one power of the velocity. As ${\cal A}_3$
describes $P$-wave quark-antiquark production it is proportional to the $l=1$
component of the Coulomb Green function at zero distance,
\begin{eqnarray}
 G^1(a,v,m,\nu) & = &
 \frac{m^4}{4\pi}\bigg\{\,
 i\,v^3 
- a\,v^2\,\left[\,\ln\left(\frac{-i\,v}{\nu}\right)
 - 1 + \ln 2+\gamma_E+\Psi\left(1\!-\!\frac{i\,a}{2\,v}\right)\,\right]
\nonumber \\ & & \qquad
+\, i\,\frac{v\,a^2}{4}
- \frac{a^3}{4}\,\left[\,\ln\left(\frac{-i\,v}{\nu}\right)
 - \frac{7}{4} + \ln 2+\gamma_E+\Psi\left(1\!-\!\frac{i\,a}{2\,v}\right)\,\right]
 \,\bigg\}
\nonumber \\ & &
 +\,\frac{m^4}{16 \pi}\,
 \left(\,\frac{1}{\epsilon}+\frac{2}{3}\,\right) 
 \left(\,v^2 a +\frac{a^3}{8}\,\right)
\,.
\label{deltaGCoulPwave}
\end{eqnarray}
The divergences and constants were determined by explicitly computing the
two-loop and four-loop potential graphs in the \ms scheme (with one and three
${\cal V}_c$ insertions respectively). As before, the result contains overall
divergences of the form $1/\epsilon$ and $v^2/\epsilon$ that are canceled by
vacuum counterterms.  The final result for the axial-vector correlator ${\cal
A}_3$ is
\begin{eqnarray}
{\cal A}_3(v,m,\nu)
 & = &
 \frac{4 \,N_c}{m^2}\, G^1(a,v,m,\nu)
\,.
\end{eqnarray}
The (fixed order) NLO and NNLO corrections to ${\cal A}_3$ are known\
\cite{Kuehn2,Penin1}. Since they contribute to $\sigma_{\rm tot}^{\gamma,Z}$
beyond NNLL order, they are not included in our analysis.

We emphasize again that the use of Eq.\ (\ref{vdefwidth}) in determining
$R^{v,a}$ does not provide a consistent treatment of the top quark width at NNLL
order.  This can be seen from the presence of $\nu$-dependent terms proportional
to $\alpha_s\Gamma_t/m_t \ln(\nu/a)$ in the imaginary part of Eq.\
(\ref{NNLLcrosssection})\ \cite{Hoang1}, and from a scheme dependent term
proportional to $\alpha_s\Gamma_t/m_t$ in the imaginary part of
Eq.~(\ref{deltaGCoulPwave}). These terms are energy independent and
parametrically of NNLL order.  Conceptually, they indicate that additional
renormalization and matching is necessary to achieve a consistent treatment of
all electroweak effects at this order. The issue of electroweak effects in the
total cross section will be addressed elsewhere.

%
\begin{figure}[t] 
\begin{center}
\vskip -0.2cm
\hspace{-0.2cm}
\epsfxsize=8cm
\leavevmode
\epsffile[90 440 540 720]{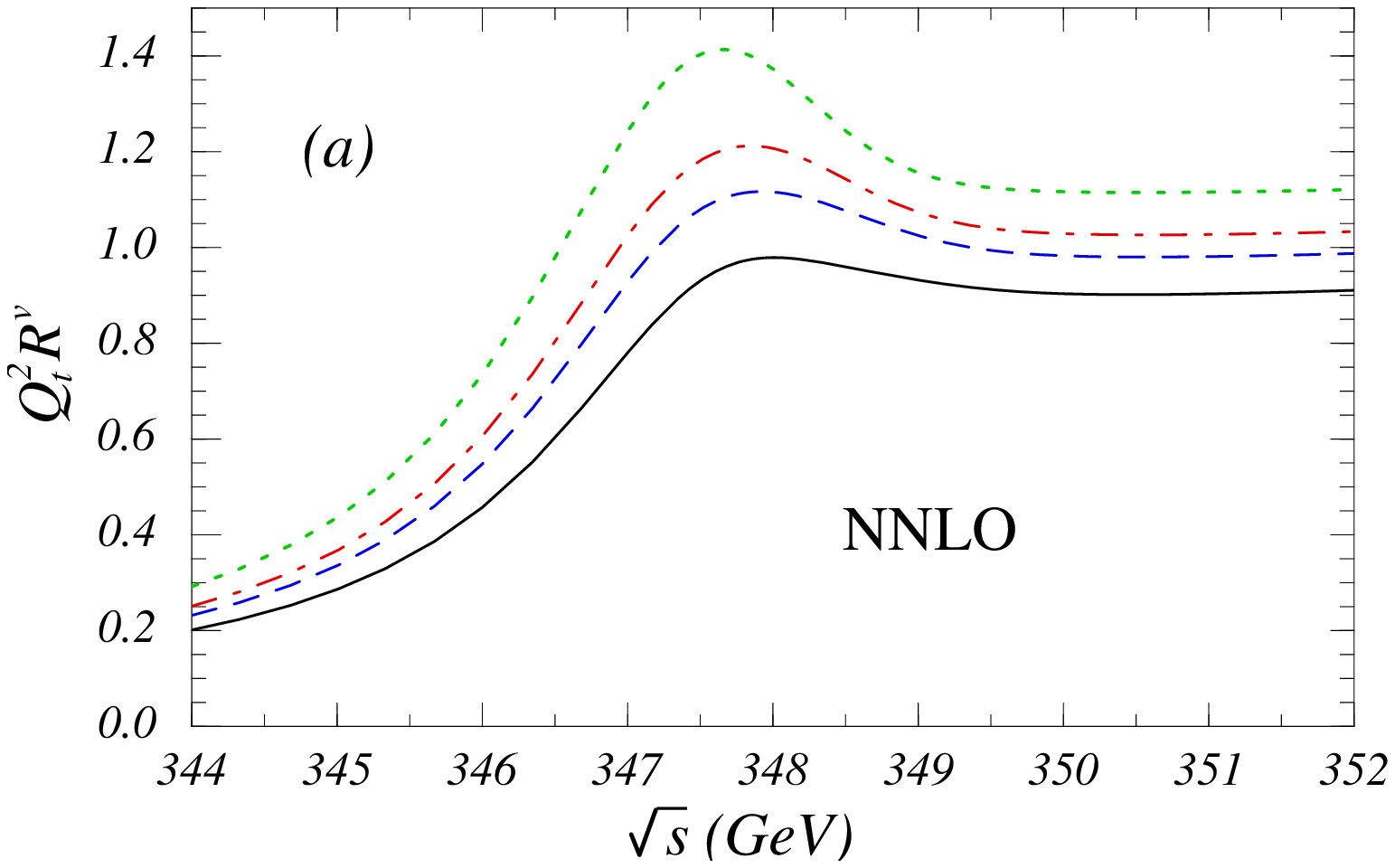}
\hspace{0.4cm}
\leavevmode
\epsfxsize=8cm
\leavevmode
\epsffile[90 440 540 720]{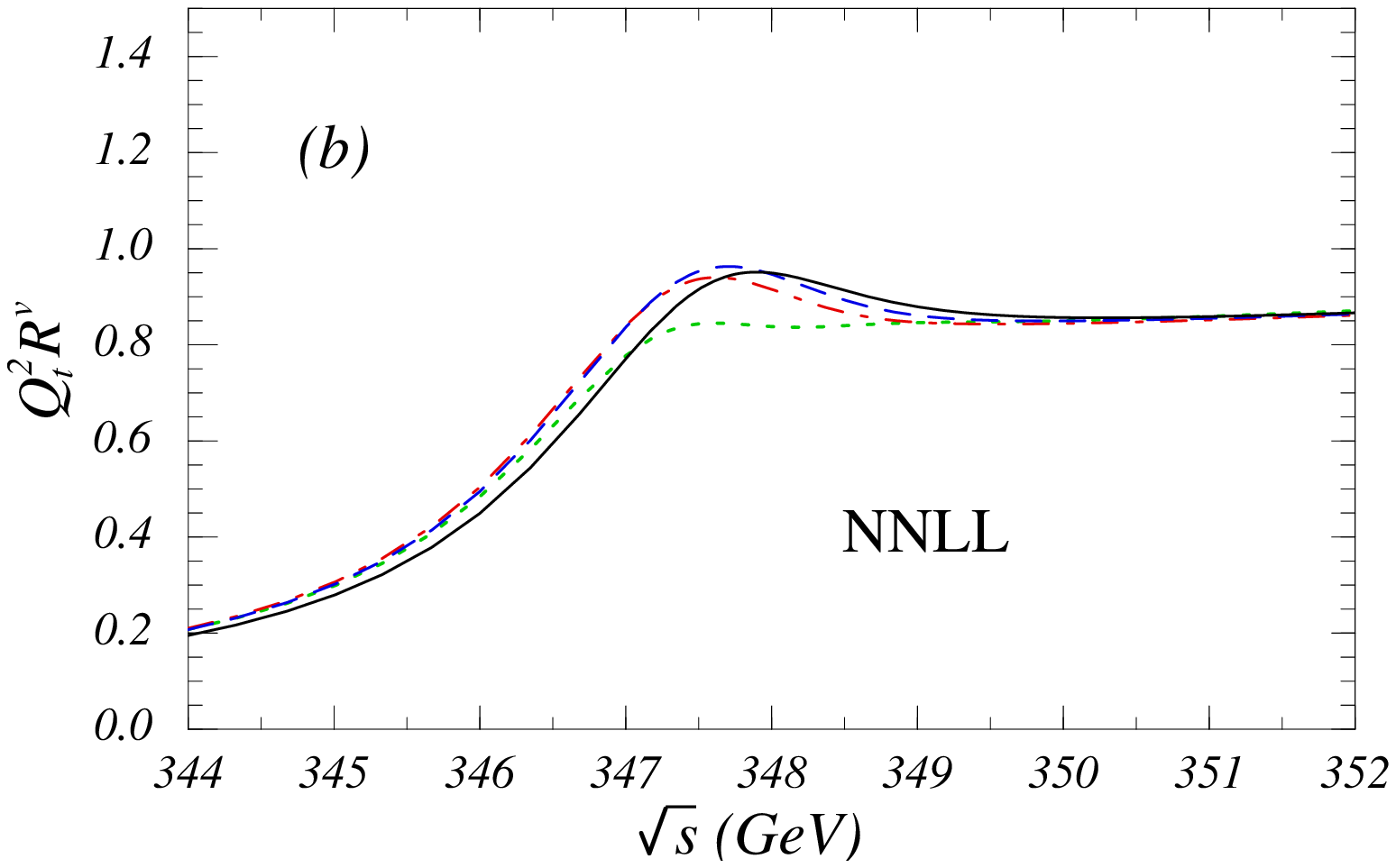}
\vskip 0.5cm 
\caption{Results for the vector current $R$-ratio in
Eq.~(\ref{Rveft}) with a fixed pole mass, $m_t=175\,{\rm GeV}$. In (a) we
reproduce the NNLO pole mass results (with matching values for Wilson
coefficients). In (b) we display our NNLL pole mass results (with running
coefficients). In both plots the dotted, dot-dashed, dashed, and solid curves
correspond to $\nu=0.1, 0.15, 0.2$, and $0.4$
respectively. \label{fig:poleplots} }
\end{center}
\end{figure}

\section{Pole mass results} \label{sectionpolemass}

Let us briefly examine the results for $R^v$ and $R^a$ in the pole mass scheme.
In Fig.~\ref{fig:poleplots}a the NNLO photon-induced cross section $Q_t^2\,R^v$
is displayed over the c.m.\ energy $\sqrt{s}$ for $\alpha_s(m_Z)=0.118$, pole
mass $m_t=175$~GeV, $\Gamma_t=1.43$~GeV, and for $\nu=0.1,0.15,0.2$ and
$0.4$. For the strong coupling four-loop running is employed and all light quark
flavors ($n_f=5$) are taken massless.  Following Ref.~\cite{Hoang3} we have
defined NNLO by taking $\nu=1$ in the current Wilson coefficients $c_{1,2}(\nu)$
and in the $\ln(-iv/\nu)$ terms in the Green functions $\delta G^{{\rm
kin},k,\delta,r}$. A lower scale $\mu_{\rm soft}=m_t\nu$ is used for factors of
$\alpha_s$ that appear in both the Green functions and in ${\cal
V}_i(\nu=1)$. At NNLO the size of the corrections and the dependence on $\nu$
are found to be similar to previous NNLO calculations~\cite{Hoang3}, indicating
that our calculational strategy does not change the fixed order results.

In Fig.~\ref{fig:poleplots}b we show the renormalization group improved NNLL
pole mass results for the same parameters and range of $\nu$ used in
Fig.~\ref{fig:poleplots}a.  From the physical point of view the appropriate
choice of the subtraction velocity parameter $\nu$ is around $C_F
\alpha_s\approx0.15$--$0.2$. With this choice all large logarithms of ratios of
the the hard and the non-relativistic scales are summed in the Wilson
coefficients, and the matrix elements of vNRQCD operators are free of large
logarithmic terms. To be conservative we show a larger range of values,
$\nu=0.1$--$0.4$, in Fig.~\ref{fig:poleplots}.  In both the NNLO and NNLL
results the dependence of the peak location on $\nu$ is a reflection of the use
of the pole mass and is considerably reduced if threshold masses are
employed~\cite{Hoang3} as we do in the next section.  From
Fig.~\ref{fig:poleplots} we already see that the dependence on $\nu$ of the
normalization of the NNLL results is significantly reduced in comparison to the
NNLO results. This is discussed in more detail in the following section.

An issue we would like to discuss here concerns the flattening of the peak that
is observed for the NNLL $\nu=0.1$ curve in Fig.~\ref{fig:poleplots}b. This
flattening turns out to be specific to the use of the pole mass together with
our calculational strategy. To see this consider the contribution from the
$n=1$, ${}^3S_1$ bound state to $R^v$. Up to trivial factors its contribution to
the cross section based on Eq.~(\ref{NNLLcrosssection}) reads
\begin{eqnarray}
R^v_{n=1} & \propto & {\rm Im}\Bigg[
\frac{|\Psi_c(0)|^2}{E_c-E-i\,\Gamma_t} +
\frac{\delta|\Psi(0)|_m^2}{E^0-E-i\,\Gamma_t} - 
\frac{\delta E_m \, |\Psi^0(0)|^2}{(E^0-E-i\,\Gamma_t)^2} \Bigg]
\,.
\label{Rvgroundstate}
\end{eqnarray}
Here $|\Psi_c(0)|^2$ and $E_c$ are the square of the wave function at the origin
and the binding energy of the ground state obtained from Eq.\
(\ref{CoulombSchroedinger}), and $\delta|\Psi(0)|_m^2$ and $\delta E_m$ are the
corresponding perturbative corrections from the potentials $\tilde V_k$, $\tilde
V_\delta$, $\tilde V_r$ and the kinetic energy correction $\bmp^4/4m^3$. The
terms $|\Psi^0(0)|^2 = m_t^3 a^3/(8\pi n^3)$ and $E^0=-m_t a^2/4$ are the LL
squared wave function at the origin and binding energy, respectively.  Explicit
formulas for $E_c$ and $\delta E_m$ including the summation of logarithms of $v$
have been presented in Ref.\ \cite{hms1}.  If the unexpanded NNLL Schr\"odinger
Equation in Eq.~(\ref{NNLLSchroedinger}) is used, then Eq.\
(\ref{Rvgroundstate}) becomes
\begin{eqnarray}
 R^v_{n=1} & \propto & {\rm Im} \Bigg[\ 
 \frac{|\Psi_c(0)|^2+\delta|\Psi(0)|_m^2}{E_c+\delta E_m-E-i\,\Gamma_t}\ \Bigg]
 \,.
\label{Rvgroundstateexact}
\end{eqnarray}
For $E\approx E_c$ and $\delta E_m>30$\,MeV the relative difference between Eq.\
(\ref{Rvgroundstateexact}) and (\ref{Rvgroundstate}) is of order $-\delta
E_m/\Gamma_t$, because $E^0-E_c\approx 1$\,GeV is large. For
$\nu=(0.1,0.15,0.2,0.3,0.4)$ and using the same parameters as in
Fig~\ref{fig:poleplots}a, we obtain $\delta E_m=(143,27,-10,-33,-40)$~MeV. Thus
for $\nu=0.1$ our calculational method leads to negative contributions of about
$10$\% at energies around the nominal peak position with respect to a
calculational strategy which instead involves the contribution in
Eq.~(\ref{Rvgroundstateexact}). This is observed in Fig.~\ref{fig:poleplots}b
for $\nu=0.1$. We have checked that one can cure this flattening by summing the
perturbative contributions in Eq.\ (\ref{Rvgroundstate}) into the single energy
denominator of Eq.\ (\ref{Rvgroundstateexact}) (see also\
\cite{Beneke1,Penin1}). However, this procedure sums terms that are subleading
according to the power counting.

Instead, we note that this effect would be small if $E^0-E_c\ll 1$~GeV. Thus, it
can be regarded as a direct consequence of the pole mass scheme where this
energy difference is large due to an infrared renormalon. In the subsequent
section we will make use of the $\SS$ mass scheme~\cite{Hoang6}, together with
our calculation strategy. In this case the flattening effect does not occur
because properly introducing the $\SS$ mass makes the analogue of the third term
in Eq.\ (\ref{Rvgroundstate}) explicitly zero.

%
\begin{figure}[t] 
\begin{center}
 \leavevmode
 \vskip -0.4cm
 \epsfxsize=8cm
 \leavevmode
 \epsffile[90 440 540 720]{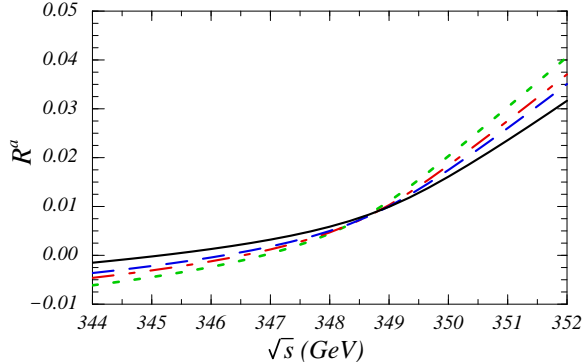}
 \vskip 0.5cm
\caption{Pole mass results for the axial-vector current $R$-ratio in
Eq.~(\ref{Rveft}). The dotted, dashed, dot-dashed, and solid curves
correspond to $\nu=0.1, 0.15, 0.2$, and $0.4$ respectively.
\label{fig:poleplots2}}
 \end{center}
\end{figure}
In Fig.~\ref{fig:poleplots2} the NNLL order axial-vector current induced cross
section $R^a$ is displayed over the c.m.\ energy $\sqrt{s}$ for $m_t=175$~GeV,
$\alpha_s(m_Z)=0.118$ and $\Gamma_t=1.43$~GeV for $\nu=0.1,0.15,0.2$ and $0.4$.
As expected from the power counting, $R^a$ is at the percent level and
suppressed by two orders of magnitude with respect to $R^v$ at c.m.\ energies
close to the $\SS$ peak and continuously grows for larger energies. No peak-like
structure is visible since the $P$-wave wavefunctions are zero at the
origin. For $\sqrt{s}\lsim 348$~GeV it is conspicuous that $R^a$ becomes
slightly negative. This unphysical behavior of $R^a$ arises because we have not
included a complete treatment of top width effects at NNLO.  In particular there
are short-distance corrections which in the non-relativistic theory correct for
the true physical top phase space~\cite{Hoang2}. Integration over the true top
phase space, of course, leads to a positive cross section. As the difference
between the approximate and true phase space integration involves only hard
momenta, it corresponds to a short-distance correction. We note that unlike
$\overline{\rm MS}$, in a hard cutoff scheme the cross section remains
positive~\cite{Hoang2} because there is no way to generate a negative result by
cutting off the integration region of a squared matrix element.  In any case,
like other known subleading width effects these corrections are quite small, of
order $\Gamma_t/m_t \sim 0.01$ for both $R^v$ and $R^a$~\cite{Hoang2}.

\section{The $\SS$ top quark mass }
\label{sectionquarkmass}

Although the heavy quark pole mass is IR-finite\ \cite{Kronfeld1,Tarrach1} and
convenient to use in bound state calculations, it suffers from a linear
sensitivity to small momenta, which leads to an ambiguity of order $\lqcd$
associated to the notion of the pole mass\ \cite{polerenormalon}. In practical
bound state calculations based on the Schr\"odinger equation this ambiguity
turns up as large $r$-independent perturbative corrections in the coordinate
space Coulomb potential $V_c(r)$. These corrections grow asymptotically like
$-\mu_S\,\alpha_s^n\,n!$~\cite{aglietti}. Because the top quark mass can be
determined from the threshold scan at a future lepton collider with experimental
uncertainties smaller than $\lqcd$\ \cite{Peralta1}, it is advantageous to
abandon the pole mass definition and replace it with a so-called ``threshold
mass''.

Threshold masses are short-distance mass definitions, i.e.\ their ambiguity is
parametrically smaller than $\lqcd$, and the expansion relating them to the
$\overline{\mbox{MS}}$ mass converges very well. They remove the large
perturbative corrections in a systematic way order by order in perturbation
theory based on the fact that the total static energy $E_{\rm stat}=2
m_t+V_c(r)$ is free of an ambiguity of order $\lqcd$\ \cite{Hoang5,Beneke3}. At
the same time, threshold masses differ from the pole mass by corrections that
are parametrically of order $m\alpha_s^2$ and comply with the non-relativistic
power-counting\ \cite{Beneke3} (see Eq.\ (\ref{NNLLSchroedinger}), where all
leading order terms are of order $m\,\alpha_s^2$). In other words, threshold
masses account for the fact that the quark is off-shell only by a small amount
of the order $v^2\sim\alpha_s^2\ll 1$.  In this work we use the $\SS$ top quark
mass definition\ \cite{Hoang2}, which is defined as half of the mass of a
perturbative contribution to the fictitious stable ${}^3S_1$ toponium ground
state.  In recent NNLO calculations of the threshold cross section alternative
threshold mass definitions have also been used such as the ``kinetic mass''\
\cite{Bigi1}, and the ``potential subtracted mass''\ \cite{Beneke3} and variants
of it\ \cite{Groote1}.  (See Ref.\ \cite{Hoang3} for a comparison of threshold
mass definitions.)

The relation between the $\SS$ mass and the pole mass can be obtained from the
${}^3S_1$ ground state ($n=1$) solution of Eq.\ (\ref{NNLLSchroedinger}) and
reads,
\begin{eqnarray}
M_t^{\mbox{\sSS}} & = &
m_t\,\Big\{\,
1 - \left[\,\Delta^{\mbox{\tiny LL}}\,\right]
  - \left[\,\Delta^{\mbox{\tiny NLL}}\,\right]
  - \left[\,\Delta_c^{\mbox{\tiny NNLL}}+\Delta_m^{\mbox{\tiny NNLL}}\,\right]
\,\Big\}
\,,
\label{1Spole}
\end{eqnarray}
where [recalling that $a\equiv -{\cal V}_c^{(s)}(\nu)/(4\pi)$]
\begin{eqnarray}
\Delta^{\mbox{\tiny LL}}(\nu) & = & \frac{a^2}{8}
\,,
\label{DeltaLL}
\\[2mm]  
 \Delta^{\mbox{\tiny NLL}}(\nu) & = & \frac{a^3}{8\pi\, C_F}\, 
 \,\bigg[\, \beta_0\,\bigg( L + 1 \,\bigg) + \frac{a_1}{2}  \,\bigg]
 \,,
\label{DeltaNLL}  \nn
\\[2mm] 
\Delta_c^{\mbox{\tiny NNLL}}(\nu) & = &
\frac{a^4}{8\pi^2\,C_F^{\,2}}\, \bigg[\,
\beta_0^2\,\bigg(\, \frac{3}{4} L^2 +  L + 
                             \frac{\zeta_3}{2} + \frac{\pi^2}{24} +
                             \frac{1}{4} 
\,\bigg) + 
\beta_0\,\frac{a_1}{2}\,\bigg(\, \frac{3}{2}\,L + 1
\,\bigg)
\nonumber\\
& & \hspace{1.5cm} +
\frac{\beta_1}{4}\,\bigg(\, L + 1
\,\bigg) +
\frac{a_1^2}{16} + \frac{a_2}{8} 
\,\bigg]
\,,
\label{DeltaNNLLc}  \nn
\\[2mm]
\Delta_m^{\mbox{\tiny NNLL}}(\nu) & = &
 - \frac{a^2}{8}\,{\cal V}_k^{(s)}(\nu)
 -\,\frac{a^3}{8\pi}\,\bigg[ \frac{{\cal V}_2^{(s)}(\nu)}{2} + 
 {\cal V}_s^{(s)}(\nu) + \frac{3 {\cal V}_r^{(s)}(\nu)}{8}\bigg]
 + \frac{5}{128}\,a^4 
\,,
\label{DeltaNNLLm}  \nn
\\[2mm] 
L & \equiv & 
\ln\Big(\frac{\nu}{a}\Big)
\,. \nn
\end{eqnarray}
For convenience, we have displayed the LL, NLL and NNLL contributions
separately. The term $\Delta_c^{\mbox{\tiny NNLL}}$ denotes the NNLL corrections
from the Coulomb potential $\tilde V_c$, and $\Delta_m^{\mbox{\tiny NNLL}}$ the
ones from the potentials $\tilde V_\delta$, $\tilde V_r$ and $\tilde V_k$ and
from the kinetic energy correction $\bmp^4/4m^3$. For $\Delta^{\rm NLL}$ and
$\Delta^{\rm NNLL}_{c,m}$ the difference between using $a$ and
$C_F\,\alpha_s(M_t^{\sSS}\nu)$ only enters beyond NNLL. Using the
non-relativistic power-counting, which is required to correctly eliminate the
pole mass, the inverse of Eq.\ (\ref{1Spole}) reads,
\begin{eqnarray}
m_t & = &
M_t^{\mbox{\sSS}}\,\Big\{\,
1 + \left[\,\Delta^{\mbox{\tiny LL}}\,\right]
  + \left[\,\Delta^{\mbox{\tiny NLL}}\,\right]
  + \left[\,(\Delta^{\mbox{\tiny LL}})^2 
+ \Delta_c^{\mbox{\tiny NNLL}}
+ \Delta_m^{\mbox{\tiny NNLL}}\,\right]
\,\Big\}
\,,
\label{pole1S}
\end{eqnarray}
where the LL, NLL and NNLL corrections are each given in brackets.  It is
mandatory to use relation~(\ref{pole1S}) at LL, NLL and NNLL for the LL, NLL and
NNLL cross section at the same low-scale subtraction velocity $\nu$ in order to
guarantee the proper cancellation of the large corrections associated with the
pole mass ambiguity.

In all of the Wilson coefficients for the potentials and currents the
replacement $m_t\to M_t^{\sSS}$ suffices at NNLL order. This is because $m_t$ and
$M_t^{\sSS}$ differ by an order $v^2$ amount. Expanding the small logarithm,
$\ln(1+\Delta_{\rm LL})$, gives
\begin{eqnarray} \label{alpsexpn}
  \alpha_s(m_t) &=& \alpha_s(M_t^{\sSS}) 
   -\frac{\beta_0}{2\pi} \alpha_s^2(M_t^{\sSS}) \Delta^{\rm LL} + \ldots \,.
\end{eqnarray}
The expansions for $\alpha_s(m_t\nu)$ and $\alpha_s(m_t\nu^2)$ in terms of
$\alpha_s(M_t^{\sSS}\nu)$ and $\alpha_s(M_t^{\sSS}\nu^2)$ are analogous.  Since
$\Delta^{\rm LL}\sim v^2$ the second term in Eq.~(\ref{alpsexpn}) gives a $v^3$
correction, and in the cross section enters at ${\rm N}^3{\rm LL}$ (or higher).

Implementing the replacement in Eq.~(\ref{pole1S}) in the low energy Green
functions is more involved. We implement the additional corrections in the
Schr\"odinger equation in Eq.~(\ref{NNLLSchroedinger}) following the
calculational strategy used in Sec.\ \ref{sectioncrosssection}.  The corrections
in Eq.\ (\ref{pole1S}) that arise from the Coulomb potential,
$\Delta^{\mbox{\tiny LL}}$, $\Delta^{\mbox{\tiny NLL}}$ and
$\Delta_c^{\mbox{\tiny NNLL}}$, are implemented through Eq.\
(\ref{CoulombSchroedinger}), which we solve exactly. This accounts for the
corrections arising from the definition of the energy $E=\sqrt{s}-2m_t$ and the
corrections coming from the leading $\bmp^2/m_t$ kinetic energy term. (For the
latter only the LL correction $\Delta^{\mbox{\tiny LL}}$ is needed.)  The term
$\Delta_m^{\mbox{\tiny NNLL}}$ from the energy definition is treated
perturbatively and leads to an additional correction to the correlator ${\cal
A}_1$ that reads,
\begin{eqnarray}
\delta {\cal A}_1(v,M_t^{\sSS},\nu)
& = &
6 \,N_c\: \delta G^{\sSS}(a,v,M_t^{\sSS},\nu)
\,,
\nonumber
\\[4mm]
\delta G^{\sSS}(a,v,M_t^{\sSS},\nu)
& = &
-\,\frac{\Delta_m^{\mbox{\tiny NNLL}}}{v}\,
\frac{d}{dv}\,G^0(a,v,M_t^{\sSS},\nu)
\,.
\label{deltaA2}
\end{eqnarray}
$\delta G^{\sSS}$ is an order $v^2$ correction since $v^{-1} d/dv\sim v^{-2}$
compensates $\Delta^{\rm NNLL}_m\sim v^4$.  Finally, in all $v^2$-suppressed
Green functions (Sec.\ \ref{sectioncrosssection}, Eqs.\
(\ref{deltaGkinetic})--(\ref{deltaGr}) and ${\cal A}_{2,3}$), the LL correction
$\Delta^{\mbox{\tiny LL}}$ needs to be included using
\begin{eqnarray}
 v & = & \left(\frac{\sqrt{s}-2M_t^{\sSS}(1+\Delta^{\mbox{\tiny LL}})+i\Gamma_t}
 {M_t^{\sSS}}\right)^{\frac{1}{2}}
\label{vdefwidthM1S}
\end{eqnarray} 
rather than Eq.\ (\ref{vdefwidth}) as the definition of the velocity, and the
pole mass $m_t$ has to be replaced by $M_t^{\sSS}$ everywhere else.  This
procedure correctly implements the $\SS$ mass scheme at NNLL order.
 
In the $\SS$ mass scheme multiple bound state energy poles in ${\cal A}_1$
caused by the perturbative treatment of the potentials $\tilde V_{\delta,r,k}$
and the kinetic energy correction only exist for higher radial excitations
($n\ge 2$), but not for the ground state ($n=1$) that is responsible for the
$\SS$ peak visible in the lineshape. This can be seen from Eq.\ (\ref{deltaA2}),
which exactly cancels the ground state double pole terms from Eqs.\
(\ref{deltaGkinetic})--(\ref{deltaGr}).  The corrections that arise from
removing the double pole terms for $n\ge 2$ by summing the
$\Delta_m^{\mbox{\tiny NNLL}}$ energy shifts into single energy denominators are
found to be at the per-mille level and negligible.

In non-threshold processes the $\overline{\mbox{MS}}$ top quark mass is
typically a more convenient mass definition to use than a threshold mass.  The
$\SS$ mass is a scale-independent quantity, and its relation to the
$\overline{\mbox{MS}}$ mass can be determined using the upsilon
expansion~\cite{Hoang6} at $\mu=m_t$ ($\nu=1$).  Explicit formul\ae\ for the
determination of the top $\overline{\mbox{MS}}$ mass from the top $\SS$ mass can
be found in Ref.~\cite{Hoang7}.

\section{Discussion}
\label{sectiondiscussion}
%
%

%
\begin{figure}[t] 
\begin{center}
 \leavevmode
\epsfxsize=3.7cm
\leavevmode
\epsffile[220 580 420 710]{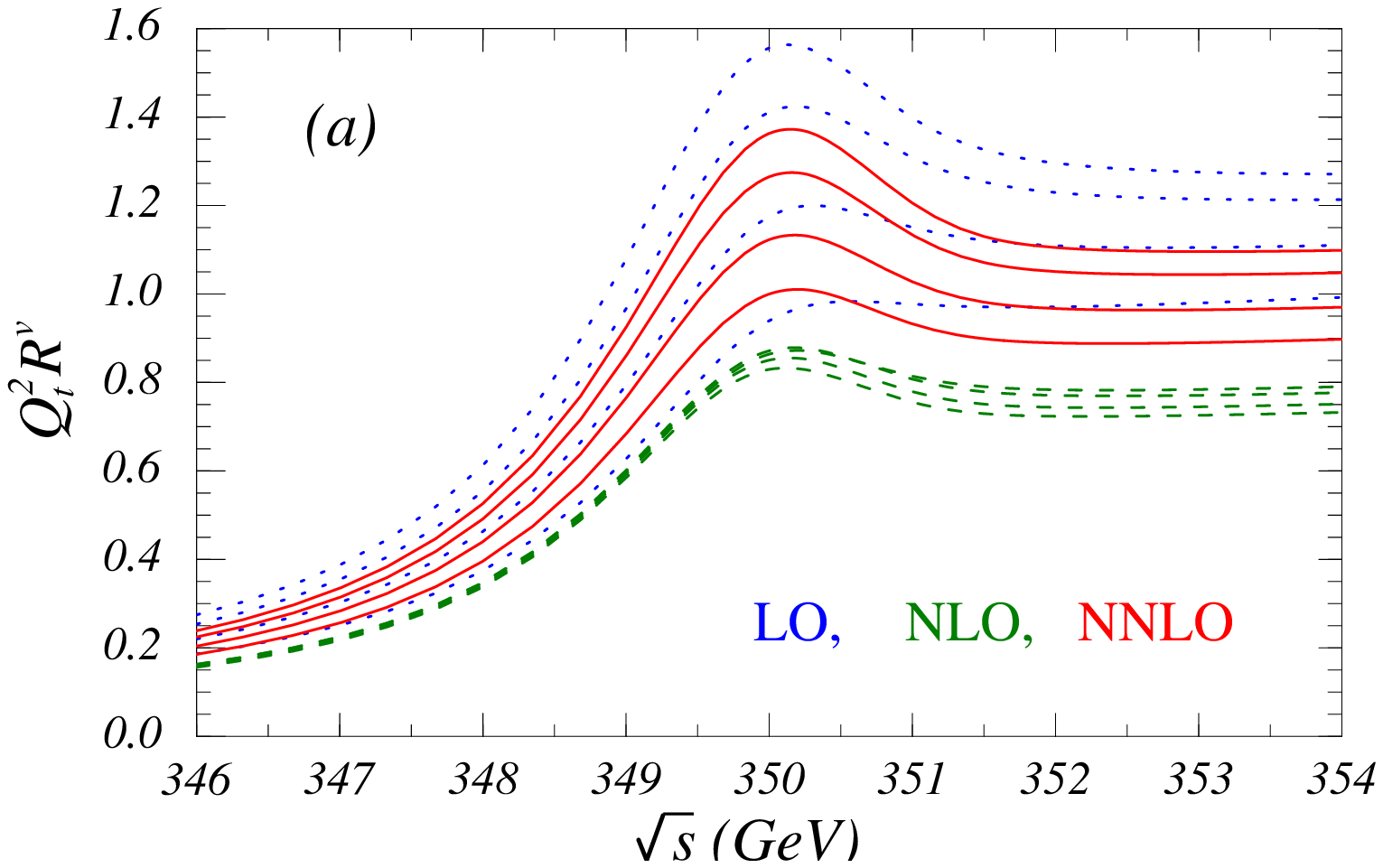}
\hspace{4.5cm}
\leavevmode
\epsfxsize=3.7cm
\leavevmode
\epsffile[220 580 420 710]{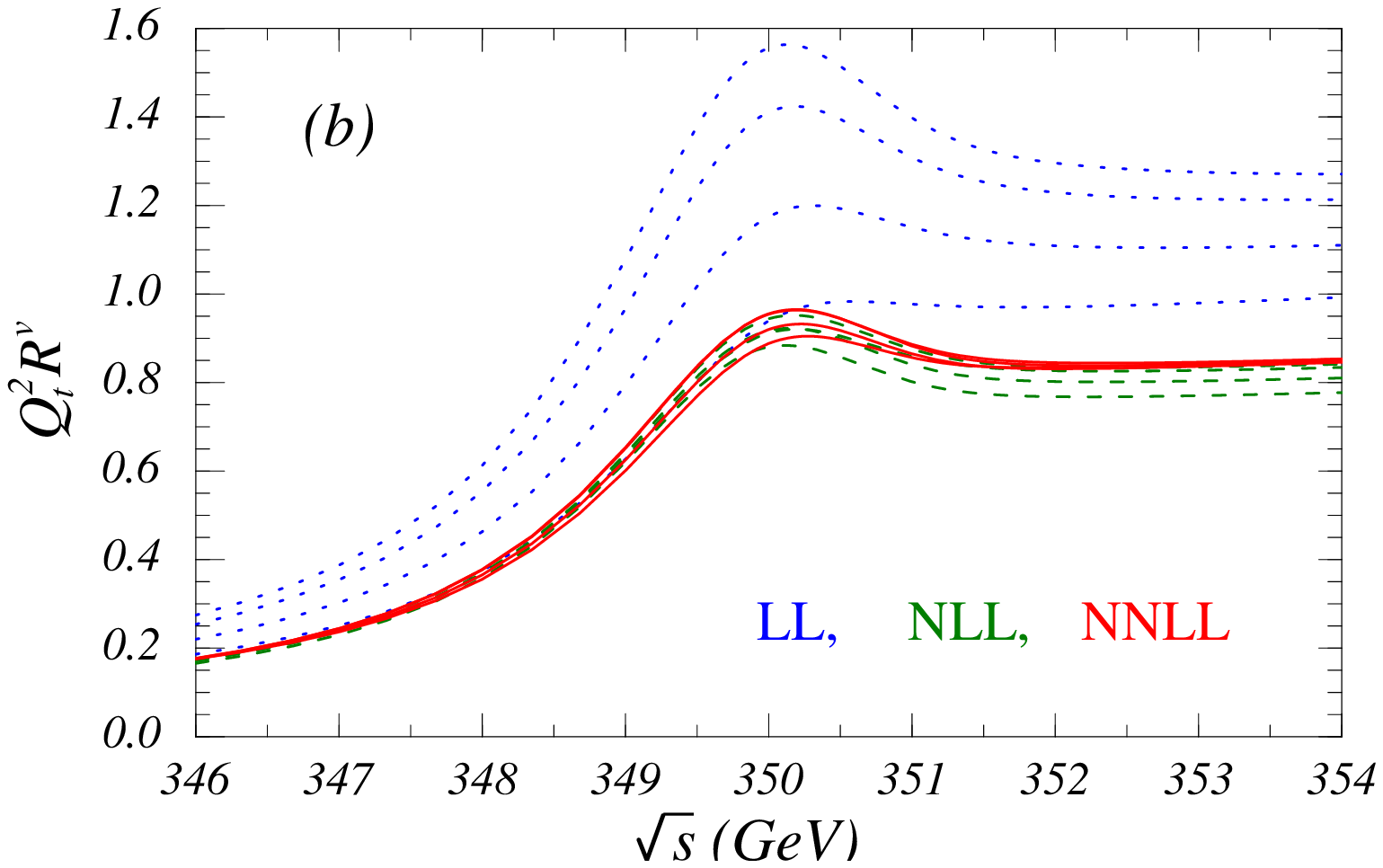}
\vskip 3.cm
\caption{Comparison of $Q_t^2 R^v$ with fixed $M_t^{\sSS}$ mass for
the fixed order and resummed expansions. The dotted, dashed, and solid
curves in a) are LO, NLO, and NNLO, and in b) are LL, NLL, and NNLL order.  For
each order four curves are plotted for $\nu=0.1$, $0.125$, $0.2$,
and $0.4$. \label{fig:m1Splots} } \end{center}
\end{figure}
\begin{table}[t!]
\begin{center}
\begin{tabular}{cll|rrrr}
 & $\sqrt{s}$ (GeV) & & $347$ & $350$ & $353$ & \\ 
 \hline
 & $Q_t^2 R^v_{\rm LL}$ & 
  $\nu=0.1\ \ $   & $0.387$ & $1.556$ & $1.276$ \\
 & \hspace{0.9cm}\raisebox{0.2cm}{} &  
  $\nu=0.125$\,\, & $0.355$  & $1.411$ & $1.215$  \\ 
 & \hspace{0.9cm}\raisebox{0.2cm}{} &  
  $\nu=0.2$       & $0.302$  & $1.175$ & $1.105$  \\ 
 & \hspace{0.9cm}\raisebox{0.2cm}{} &  
  $\nu=0.275$       & $0.276$  & $1.054$ & $1.043$  \\ 
 & \hspace{0.9cm}\raisebox{0.2cm}{} &  
  $\nu=0.4$       & $0.251$  & $0.940$ & $0.980$  \\ \hline
 & $Q_t^2 R^v_{\rm NLL}$ & 
  $\nu=0.1\ \ $   & $0.230$ & $0.881$ & $0.770$ \\
 & \hspace{0.9cm}\raisebox{0.2cm}{} &  
  $\nu=0.125$     & $0.237$  & $0.917$ & $0.804$  \\ 
 & \hspace{0.9cm}\raisebox{0.2cm}{} &  
  $\nu=0.2$       & $0.243$  & $0.944$ & $0.835$  \\ 
 & \hspace{0.9cm}\raisebox{0.2cm}{} &  
  $\nu=0.275$       & $0.242$  & $0.937$ & $0.837$  \\ 
 & \hspace{0.9cm}\raisebox{0.2cm}{} &  
  $\nu=0.4$       & $0.237$  & $0.912$ & $0.827$  \\ \hline
 & $Q_t^2 R^v_{\rm NNLL}$ & 
  $\nu=0.1\ \ $   & $0.237$ & $0.888$ & $0.842$ \\
 & \hspace{0.9cm}\raisebox{0.2cm}{} &  
  $\nu=0.125$     & $0.240$  & $0.920$ & $0.836$  \\ 
 & \hspace{0.9cm}\raisebox{0.2cm}{} &  
  $\nu=0.2$       & $0.244$  & $0.955$ & $0.841$ \\
 & \hspace{0.9cm}\raisebox{0.2cm}{} &  
  $\nu=0.275$       & $0.245$  & $0.961$ & $0.845$  \\ 
 & \hspace{0.9cm}\raisebox{0.2cm}{} &  
  $\nu=0.4$       & $0.244$  & $0.955$ & $0.846$ 
\end{tabular}
\end{center}
{\tighten \caption{Numerical values of $Q_t^2 R^v$ which appear in the NNLL
results in Fig.~\ref{fig:m1Splots}b.}
\label{tab:m1Svalues} }
\end{table}

In this section we carry out a detailed analysis of $R^v$ and $R^a$ in the $\SS$
mass scheme with the main emphasis on assessing the remaining theoretical
uncertainties in our computation.  In Fig.\ \ref{fig:m1Splots} we have displayed
results for $Q_t^2 R^v$ over the c.m.\ energy $\sqrt{s}$ for
$M_t^{\sSS}=175$~GeV, $\alpha_s(m_Z)=0.118$ and $\Gamma_t=1.43$~GeV. For the
strong coupling four-loop running is employed and all light quark flavors
($n_f=5$) are taken massless.  Fig.~\ref{fig:m1Splots}a shows results at LO
(dotted blue lines), NLO (dashed green lines) and NNLO (solid red lines), while
Fig.~\ref{fig:m1Splots}b shows the resummed results at LL (dotted blue lines),
NLL (dashed green lines) and NNLL (solid red lines) order.  At each order four
curves are displayed corresponding to $\nu=0.1$, $0.125$, $0.2$, and $0.4$. At
$\sqrt{s}=350$~GeV the LO, LL, and NNLO upper through lower lines correspond
monotonically to $\nu=0.1$ through $\nu=0.4$. In contrast, the upper through
lower NLO lines correspond to $\nu=0.4$ through $\nu=0.1$.  At NLL the lower and
upper lines at $\sqrt{s}=350$~GeV correspond to $\nu=0.1$ and $\nu=0.2$ (whereas
$\nu=0.4$ lies in between). Finally, at NNLL the lower line is $\nu=0.1$, while
the upper line includes both $\nu=0.2$ and $\nu=0.4$ at the displayed
resolution.  Our NNLO results agree well with previous NNLO analyses, see Ref.\
\cite{Hoang3} for a synopsis of previous NNLO results. In Table\
\ref{tab:m1Svalues} we have summarized the NNLL values for $Q_t^2 R^v$ for
$\sqrt{s}=347$, $350$ and $353$~GeV with the same parameters as used in the
figure.  The values $\nu=0.125$ and $\nu=0.275$ are included in the Table since
at $\sqrt{s}=353$ the value of $Q_t^2 R^v$ is minimized near $\nu=0.125$, while
for $\sqrt{s}=350$ its value is maximized near $\nu=0.275$.

For $R^v$ the LL cross section is equivalent to the LO one because the one-loop
anomalous dimension of $c_1$ vanishes. The NLL cross section is obtained by
eliminating from Eq.~(\ref{NNLLcrosssection}) the corrections from the
$1/m$-suppressed potentials, the kinetic energy term
$({{\mbox{\boldmath $p$}}^4}/{4 m^3})$ and the operators $\O{p}{2}$ and
$\O{p}{3}$. For $c_1$ the NLL evolution and the one-loop matching condition is
employed. For the Coulomb potential, the coupling ${\cal V}_c(\nu)$ is set equal
to $-4\pi C_F \alpha_s(M_t^{\sSS}\nu)$, and only the one-loop corrections
proportional to $[\alpha_s(M_t^{\sSS}\nu)]^2$ have to be taken into account.

%
\begin{figure}[t] 
\begin{center}
\leavevmode
\epsfxsize=3.7cm
\leavevmode
\epsffile[220 580 420 710]{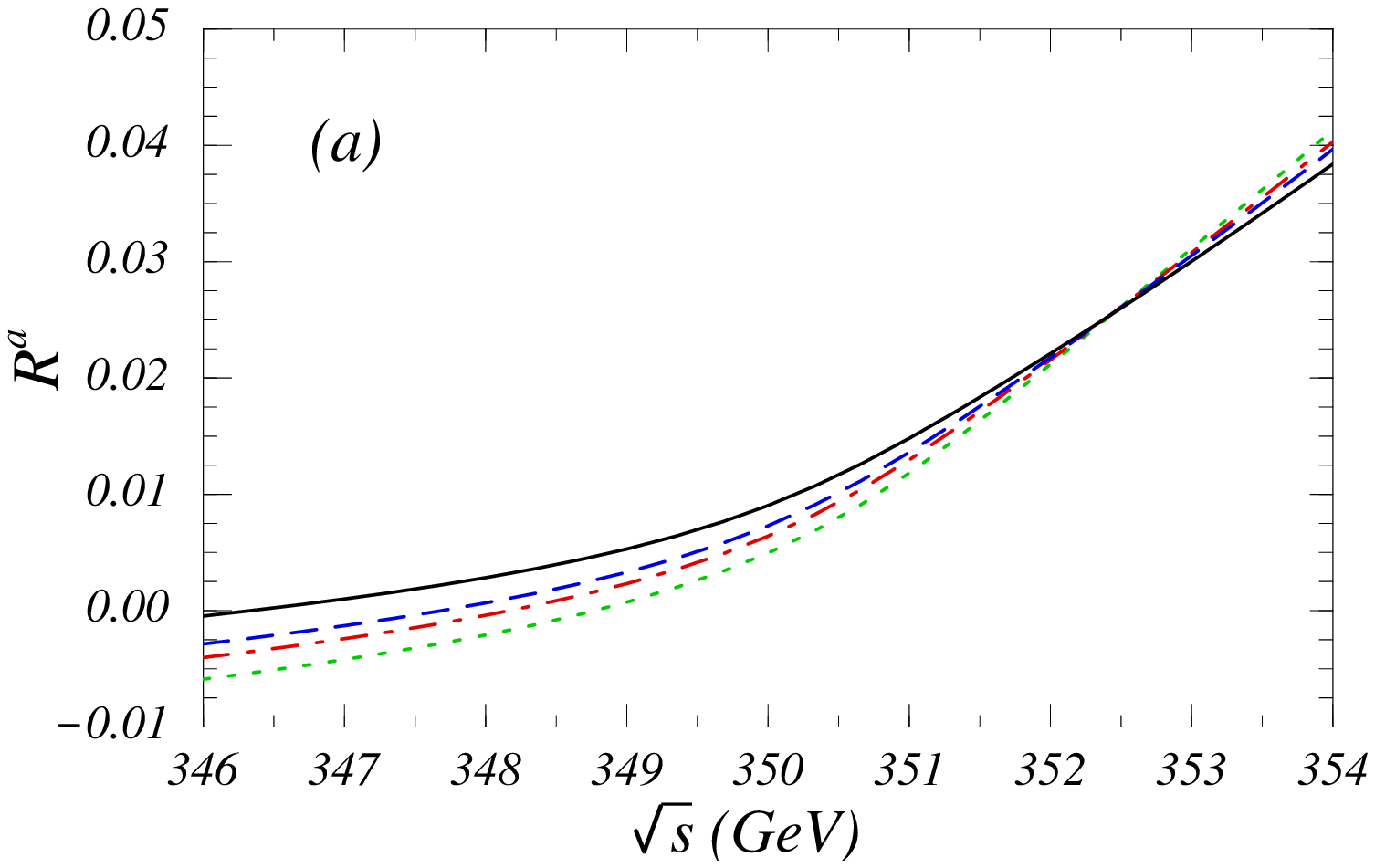}
\hspace{4.5cm}
\leavevmode
\epsfxsize=3.7cm
\leavevmode
\epsffile[220 580 420 710]{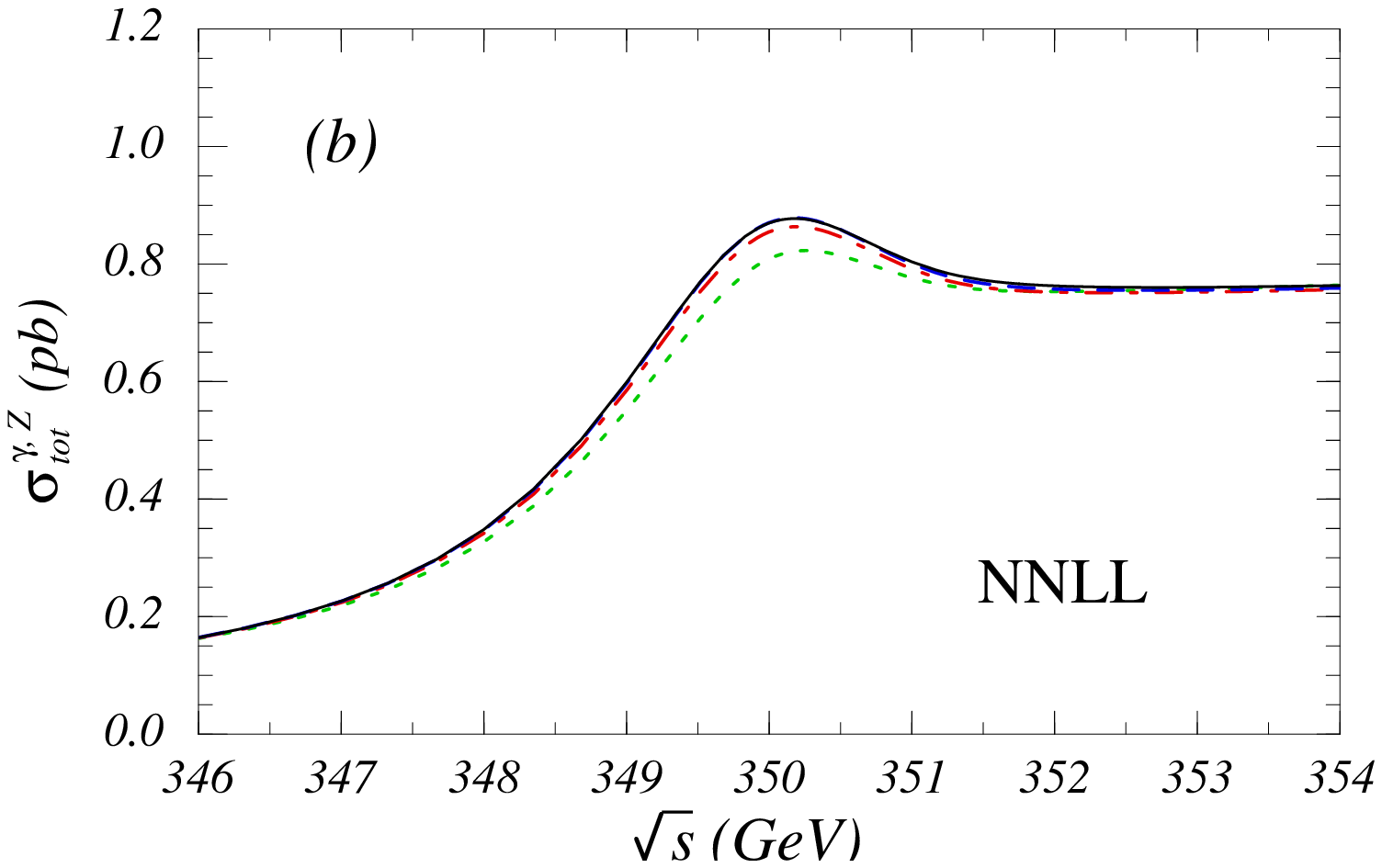}
\vskip 3.cm
 \caption{Using the $M_t^{\sSS}$ mass  a) shows the LL result
for $R^a$, and b) shows the total NNLL cross section in
$pb$. The dotted, dashed, dot-dashed, and solid curves correspond to $\nu=0.1$,
$0.15$, $0.2$, and $0.4$.
\label{fig:m1Splots2} 
}
\end{center}
\end{figure}
In Fig.\ \ref{fig:m1Splots2}a we have displayed $R^a$ at LL for
$M_t^{\sSS}=175$~GeV, $\alpha_s(m_Z)=0.118$ and $\Gamma_t=1.43$~GeV and
$\nu=0.1$, $0.15$, $0.2$, $0.4$. The lower through upper lines for
$\sqrt{s}<352$~GeV correspond monotonically to $\nu=0.1$ through
$\nu=0.4$. With the $\SS$ mass, $R^a$ dips below zero for $\sqrt{s}\lesssim
347\,{\rm GeV}$. This dip has the same origin as for the pole mass and is
discussed in section~\ref{sectionpolemass}.  Finally, in Fig.\
\ref{fig:m1Splots2}b we have displayed our total NNLL result for the $\gamma+Z$
induced cross section, using the same parameters as in Fig.\
\ref{fig:m1Splots2}a.

Comparing Figs.~\ref{fig:m1Splots}a and \ref{fig:m1Splots}b the variation of the
normalization of $R^v$ with $\nu$ at NNLL order is considerably reduced in
comparison to the NNLO result.  Equally important, the size of the NNLL
corrections is considerably smaller than the size of the corresponding NNLO
corrections.  The latter indicates that at the low scale the perturbative
corrections to the threshold $t\bar t$ cross section are converging.  In
$\sigma_{\rm tot}^{\gamma,Z}$ the remaining variation of $R^v$ with $\nu$ is
still much larger than the variation of $R^a$. Therefore, to estimate the
remaining overall uncertainty in $\sigma_{\rm tot}^{\gamma,Z}$ the variation of
$R^a$ can be neglected, and for the rest of this section we discuss only $R^v$.

%
\begin{figure}[t] 
\begin{center}
\leavevmode
\epsfxsize=3.7cm
\leavevmode
\epsffile[220 580 420 710]{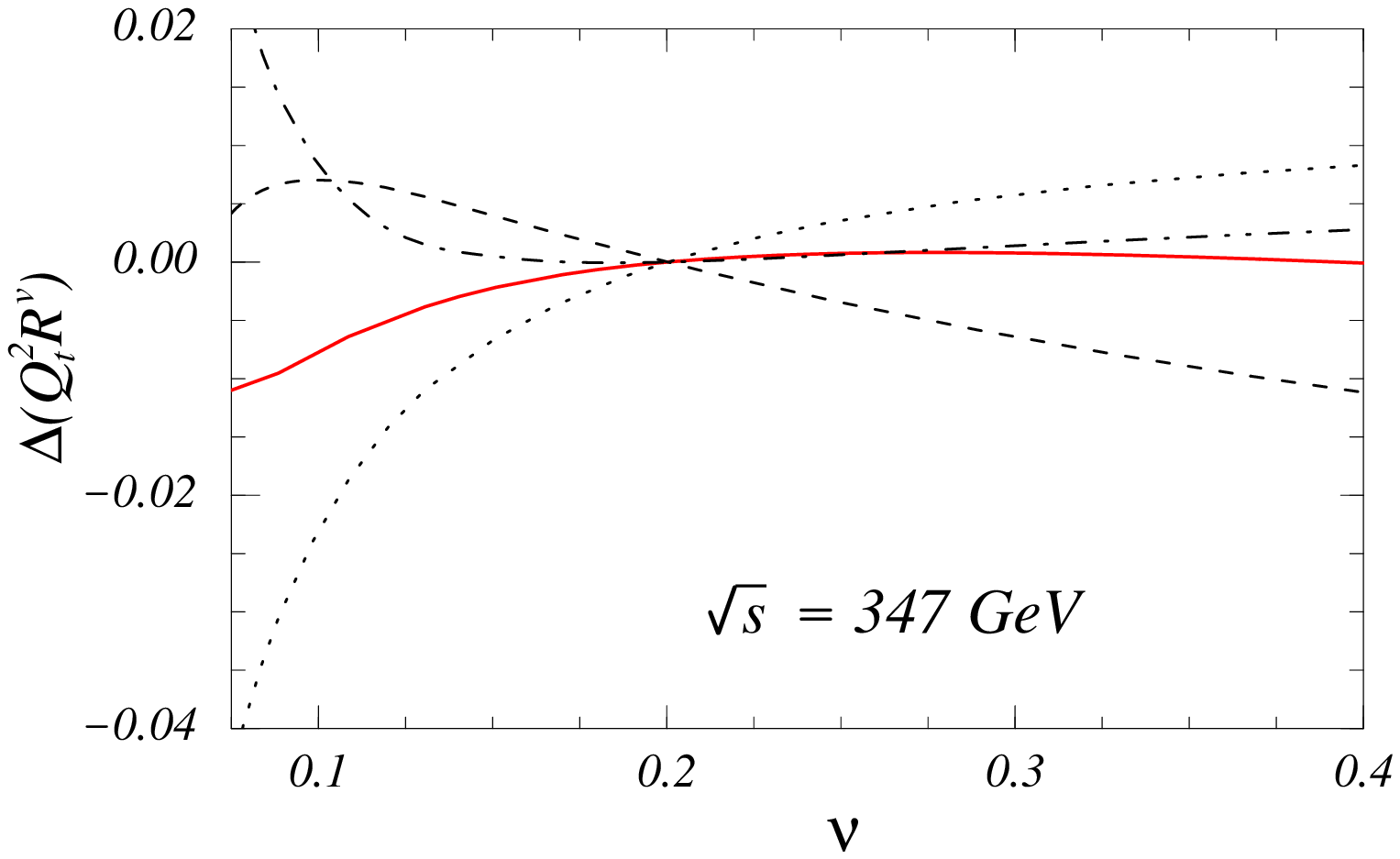}
\hspace{4.5cm}
\leavevmode
\epsfxsize=3.7cm
\leavevmode
\epsffile[220 580 420 710]{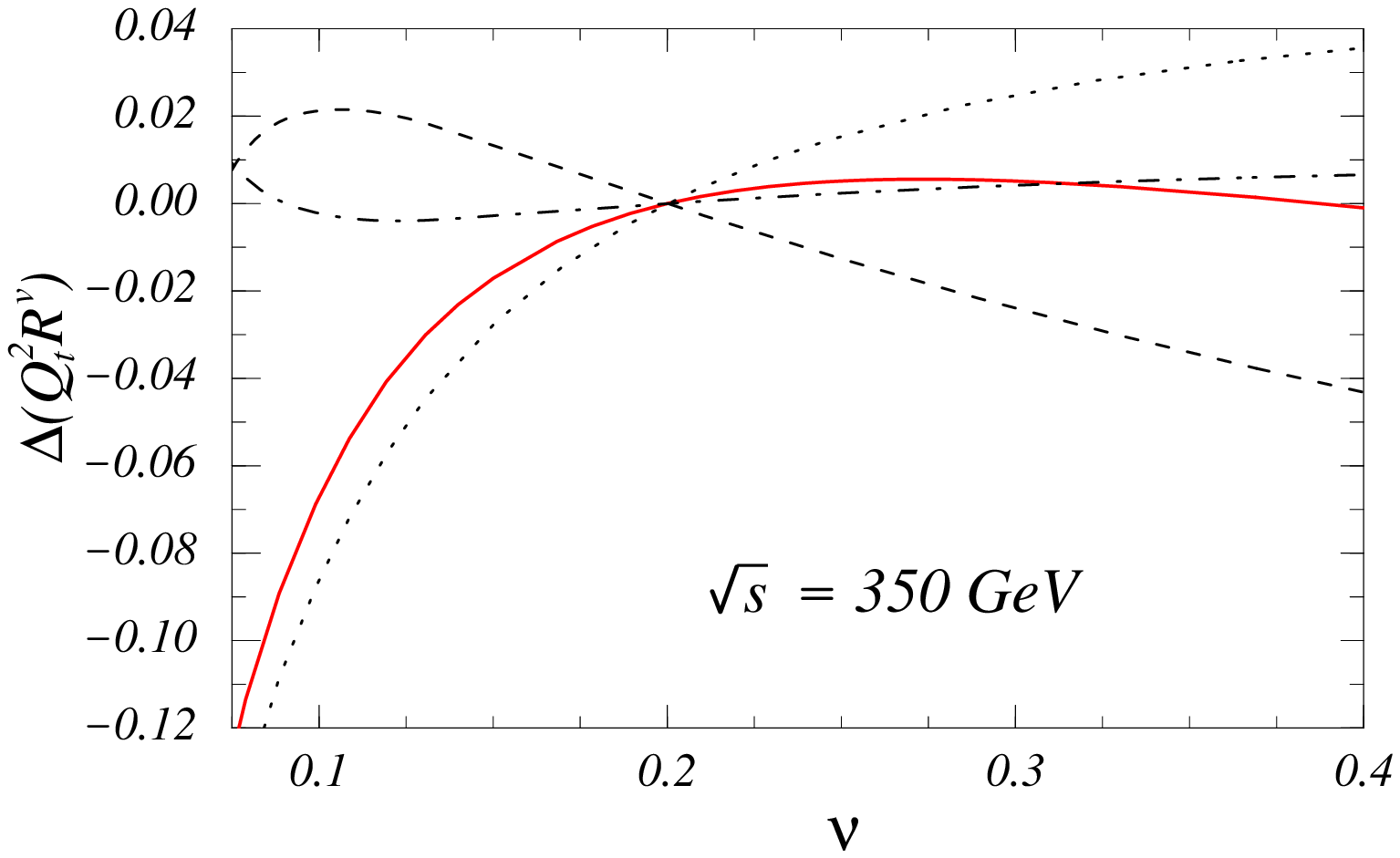}
\leavevmode
\\[3.cm]
\epsfxsize=3.7cm
\leavevmode
\epsffile[220 580 420 710]{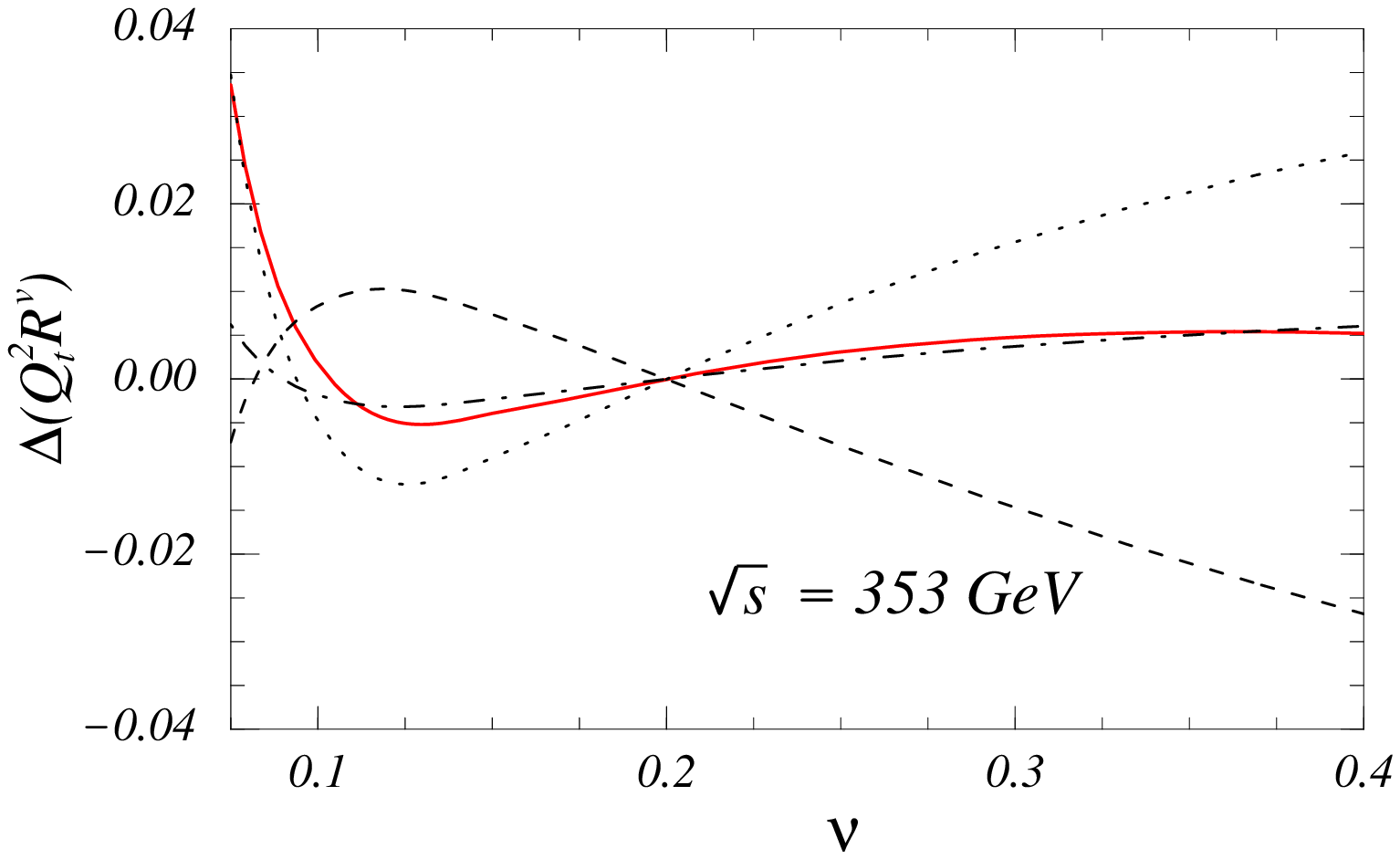}
\vskip 3.cm
 \caption{Relative scale dependence of various NNLL contributions to
$Q_t^2 R^v$. The contributions in Eq.~(\ref{NNLLcrosssection}) were divided into
those from $G^c_{v,m,\nu}$ (dashed lines), $\delta G^k$ (dotted lines), and the
sum of $\delta G^{\delta,r,{\rm kin},\sSS}$ (dot-dashed lines), while the solid
lines denote the sum of these terms.
\label{fig:vvarplots} 
}
 \end{center}
\end{figure}
First we wish to explore the origin of the reduced $\nu$ dependence.  In Fig.\
\ref{fig:vvarplots}a,b,c $Q_t^2 R^v$ (solid red line) is displayed as a function
of $\nu$ for $\sqrt{s}=347$, $350$, $353$~GeV, using the choice of parameters of
Fig.\ \ref{fig:m1Splots}. We have also displayed separately, the contributions
to $Q_t^2 R^v$ coming from $G^c$ (dashed line), $\delta G^k$ (dotted line) and
the sum of $\delta G^{\delta,r,{\rm kin},\sSS}$ (dot-dashed line). For all
curves in Fig.\ \ref{fig:vvarplots} the respective values at $\nu=0.2$ have been
subtracted. The results show that for any c.m.\ energy the $\nu$-variation of
the individual contributions to $R^v$ is much stronger than the $\nu$-variation
of the sum. As an example, for $\sqrt{s}=350$ and $\nu>0.15$, one finds that the
variation of the contributions from $G^c$ and $\delta G^k$ cancels almost
entirely, whereas the contribution from the sum $\delta G^{\delta,r,{\rm
kin},\sSS}$ is nearly $\nu$-independent. On the other hand, for $\sqrt{s}=350$
and $\nu<0.2$ the $\nu$-variation is dominated by the contribution from $\delta
G^k$ which rapidly decreases, whereas the contributions from $G^c$ and $\delta
G^{\delta,r,{\rm kin},\sSS}$ are small. The situation is quite similar for other
energies, and results directly from the evolution equation of the Wilson
coefficients. These evolution equations constitute a system of coupled
differential equations, where the values of the Wilson coefficients become
related to each other for $\nu<1$. This is different from the fixed order NNLO
calculations carried out in previous
works~\cite{Hoang1,Melnikov1,Yakovlev1,Beneke1,Nagano1,Hoang2,Penin1}, where the
proper anomalous dimensions of the potentials and currents were not taken into
account, and where the Wilson coefficients at the soft scale were obtained from
simply evaluating their hard matching conditions at the soft scale.  In
particular, in the fixed order approach, there is no numerical cancellation
between the scale-dependence that arises from the corrections to the Coulomb
potential $\tilde V_c$ and the corrections from the potential $\tilde V_k$ (see
Eqs.\ (\ref{VCoulomb}) and (\ref{Vkdr})). In addition, in the fixed order
approach, the contribution from $\tilde V_k$ becomes quite large at low scales
because it is proportional to the square of the strong coupling. In fact, in
previous fixed order NNLO calculations $\tilde V_k$ was identified as one of the
major sources of the large positive corrections to the normalization of the
total cross section. Using the VRG, on the other hand, the Wilson coefficient
for $\tilde V_k$ decreases by almost an order of magnitude between $\nu=1$ and
$\nu\sim 0.2$, and even vanishes for $\nu\approx 0.17$~\cite{amis3}. This is one
of the main reasons for the small size of the NNLL corrections.  From the
previous considerations we conclude that the cancellation of the
$\nu$-dependence and the size of the corrections that arises in the sum the
various contributions to $R^v$ is a genuine and systematic QCD effect, and, in
particular, that it is justified to take the overall $\nu$-variation of $R^v$ as
a tool to estimate the remaining theoretical uncertainties.

%
\begin{figure}[t] 
\begin{center}
 \leavevmode
 \epsfxsize=3.7cm
 \leavevmode
 \epsffile[220 580 420 710]{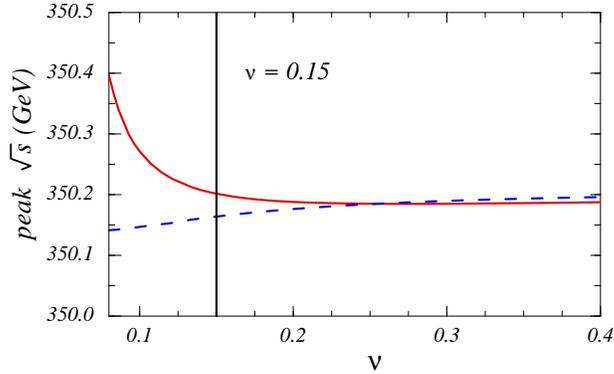}
 \vskip 2.5cm
 \caption{Position of the peak in the cross-section versus $\nu$ at NNLO
(dashed) and NNLL (solid). The vertical line at $\nu=0.15$ is a physically
motivated endpoint for the running. \label{fig:peakpos} 
}
 \end{center}
\end{figure}
Let us now consider the range of $\nu$ that should be considered for a suitable
error estimate.  As mentioned before, from the physical point of view the
appropriate choice of the subtraction velocity parameter $\nu$ is around
$\alpha_s\approx0.15$--$0.2$. With this choice all large logarithms of ratios of
the the hard scale $m_t$ and the non-relativistic scales $m_t v$ and $m_t v^2$
are summed into the Wilson coefficients, whereas the matrix elements of vNRQCD
operators are free of such logarithmic terms.  From Fig.~\ref{fig:m1Splots2}b we
see that if the lower limit for $\nu$ is taken to be the endpoint of the
running, $\nu\simeq 0.15$, then there is really only a very small uncertainty
from the residual scale dependence. To be conservative we feel that the
$\nu=0.1$ curve should be included.  The curves in Fig.\ \ref{fig:vvarplots}
show that our results become unstable for $\nu\lsim 0.1$.  This is a consequence
of the dependence of the VRG equations on the ultrasoft scale $\mu_U=m_t\nu^2$,
which becomes of the order $1$~GeV for $\nu\lsim 0.1$.  Thus, for $\nu \lsim
0.1$ the perturbation expansion is apparently unreliable.  However, this is just
an artifact of choosing $m_t \nu^2$ much smaller than the physical energy scales
in $\bar tt$, so that $\alpha_s(m_t \nu^2)$ is large.  The same feature can also
be observed in Fig.\ \ref{fig:peakpos}, where the position of the $\SS$ peak is
displayed as a function of $\nu$ for the parameter choice used previously. For
$\nu\ge 0.15$ the peak position is stable, while for $\nu\lsim 0.1$ the peak
position increases rapidly to higher energies. This is in contrast to the
expectations in the $\SS$ mass scheme, where the peak position should be stable
by construction. The behavior of the peak position for $\nu\lsim 0.1$ is a
consequence of our perturbative treatment of the corrections that arise from the
$1/m$-suppressed potentials and the kinetic energy correction. For $\nu\lsim
0.1$ these corrections become large and render the perturbative treatment
unreliable. Thus, we conclude that the range $\nu\lsim 0.1$ should not be used
for the estimate of the remaining theoretical uncertainties.

As far as values of $\nu$ larger than 0.2 are concerned the curves in Figs.\
\ref{fig:m1Splots}--\ref{fig:peakpos} show that our
results are remarkably stable up to quite large values of the subtraction
velocity, even beyond $\nu=0.4$. For the estimate of the remaining theoretical
uncertainty of the total cross section it is therefore not mandatory to fix any
strict upper bound of $\nu$. In Table.\ \ref{tab:m1Svalues} we have shown
results for $\nu=0.1$, $0.125$, $0.2$, $0.275$, and $0.4$. We consider the small
value, $\nu=0.1$, as a conservative lower bound, which is set by physical
considerations, and below which no sensible perturbative treatment is
possible. A less conservative approach would have been to take the lower bound
to be set by the typical top quark velocity, $\nu\simeq 0.15-0.2$.  The value
$\nu=0.275$, on the other hand, has been included because for this choice the
peak cross section reaches its maximum as a function of $\nu$.  We take the size
of the resulting $\nu$-variation of the cross section at the $\SS$ peak,
\begin{eqnarray}
 {\delta\sigma_{t\bar t} \over \sigma_{t\bar t} }\ =\ \pm\, 3\%
\end{eqnarray}
as the remaining uncertainty inherent to our renormalization group improved
calculation. This is an order of magnitude smaller than the uncertainty that was
associated to previous fixed order NNLO QCD calculations of the cross section.
Note that if $\nu=0.125$ was instead used as the lower bound for the running
then the $\nu$-variation is ${\delta\sigma_{t\bar t}}/{\sigma_{t\bar t} } =\pm\,
1.5\%$. In fact, the small $\nu$-dependence and the small size of the NNLL order
corrections observed in our calculation of the cross section comply with the
expectations from power counting and the natural notion that the
non-relativistic $t\bar t$ dynamics should be calculable to a high degree of
precision\ \cite{Kuehn1,Fadin1}.

It is instructive to compare the error assigned to the cross section based on
the $\nu$-variation with the size of corrections from terms that are not yet
included in our NNLL result.  Firstly, at NNLL the complete running of
$c_1(\nu)$ was not included because the three loop anomalous dimension for
$c_1(\nu)$ is not yet completely known. (We recall that already at NLL order
$[c_1(\nu\approx \alpha_s)]^2$ contains the sizeable negative (fixed order)
${\rm N}^3{\rm LO}$ normalization correction proportional to $\alpha_s^3
\ln^2\alpha_s$ that was determined in Ref.\ \cite{Kniehl1}, see Ref.\
\cite{amis3}.) Contributions to the three loop anomalous dimension are discussed
in Appendix~\ref{App_run}, and based on dimensional analysis and the known
contributions to this anomalous dimension we estimate that the uncertainty in
neglecting it is $\lesssim 2\%$.  It is also useful to consider the size of some
corrections from beyond NNLL that can be determined easily or have been obtained
earlier. We emphasize that all such corrections belong to matrix elements of the
effective theory, i.e. it is crucial to employ the proper choice for the
renormalization scale $\mu_S$ or $\mu_U$ depending on whether the corrections
come from the soft (or potential) or ultrasoft momentum regime. Let us first
consider the corrections arising from two insertions of $\tilde V_\delta$ (at
second order time-independent perturbation theory). As the correction arises
from potential momenta the soft renormalization scale has to be used. We find
that the relative corrections are at most $2.8$\% (for $\nu=0.1$ and close to
the peak). In Ref.\ \cite{Kniehl2} the corrections to the square of the $S$-wave
quarkonium wavefunction at the origin arising from the emission and reabsorption
of an ultrasoft gluon were determined.  Naturally the ultrasoft scale has to be
used in this case. These corrections can be taken as an estimate for the
ultrasoft corrections to the normalization of the total cross section. For the
ground state ($n=1$) one finds that this ultrasoft correction amounts to about
$2$\% for $\nu\approx \alpha_s$. This is also consistent with our error
estimate.

\section{Phenomenological Consequences}
\label{sectionphenomenology}
%
%
%
\begin{figure}[t] 
\begin{center}
\leavevmode
\epsfxsize=3.7cm
\leavevmode
\epsffile[220 580 420 710]{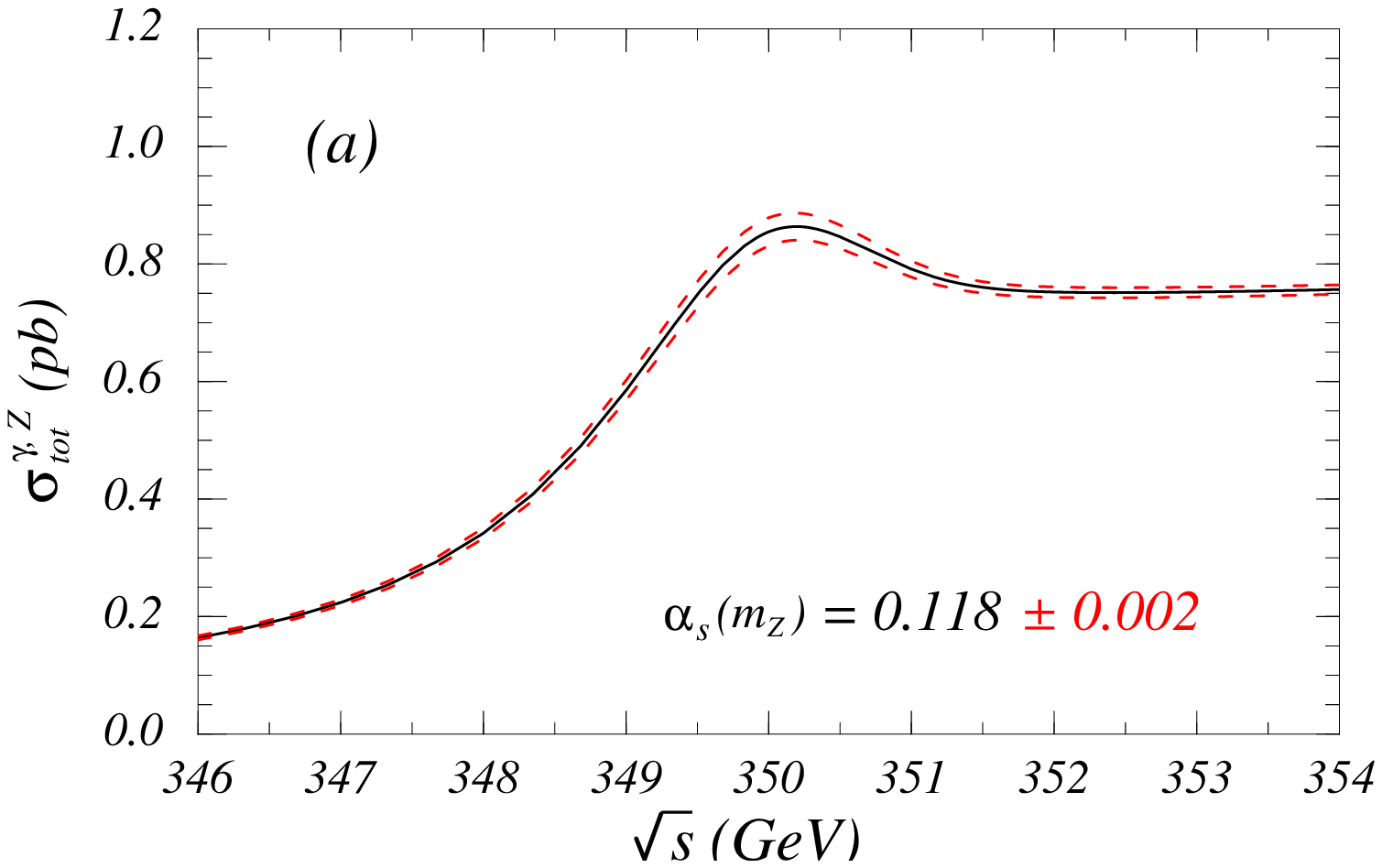}
\mbox{\hspace{4.4cm}}
\leavevmode
\epsfxsize=3.7cm
\leavevmode
\epsffile[220 580 420 710]{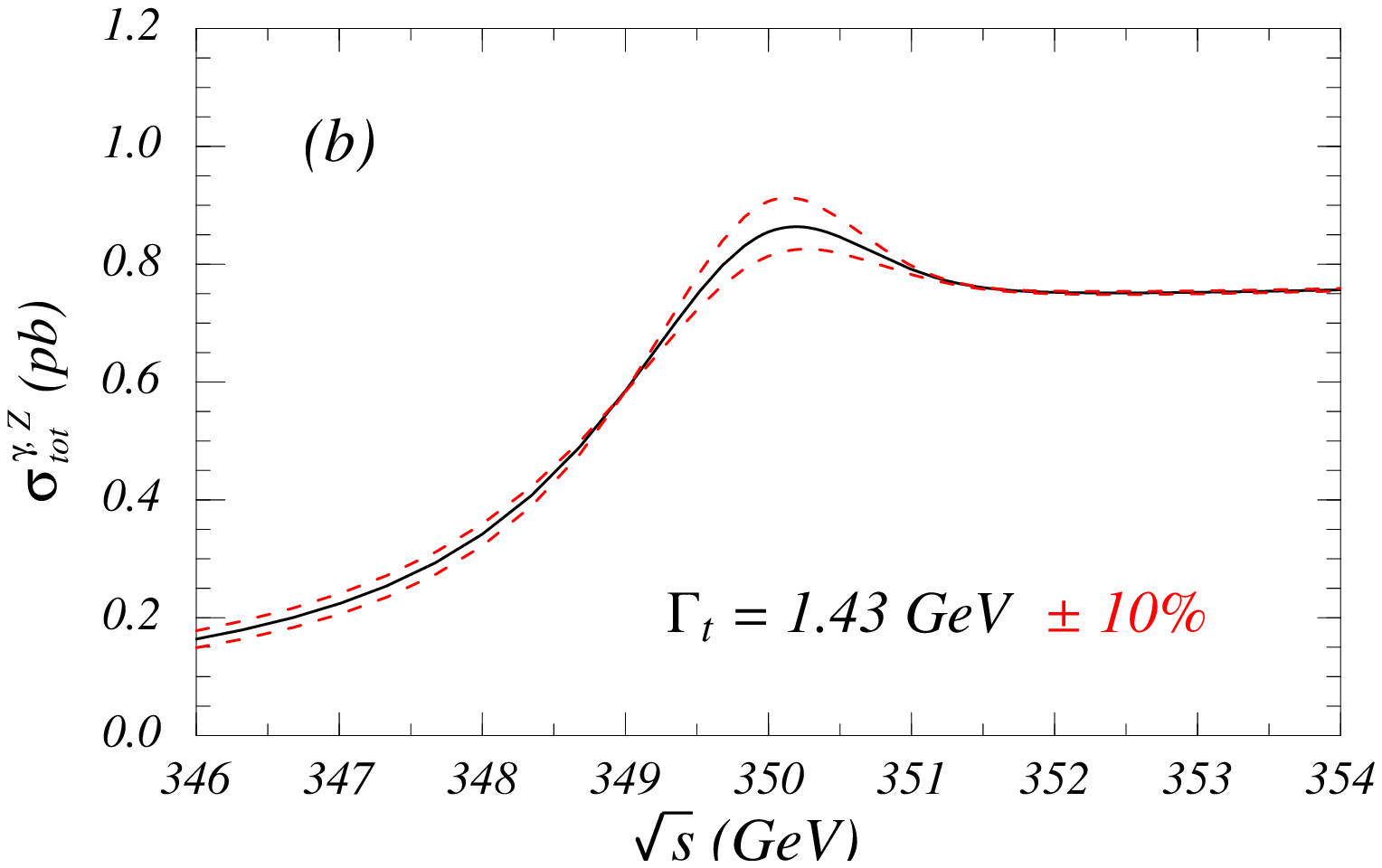}
\\[3.2cm]
\leavevmode
\epsfxsize=3.7cm
\leavevmode
\epsffile[220 580 420 710]{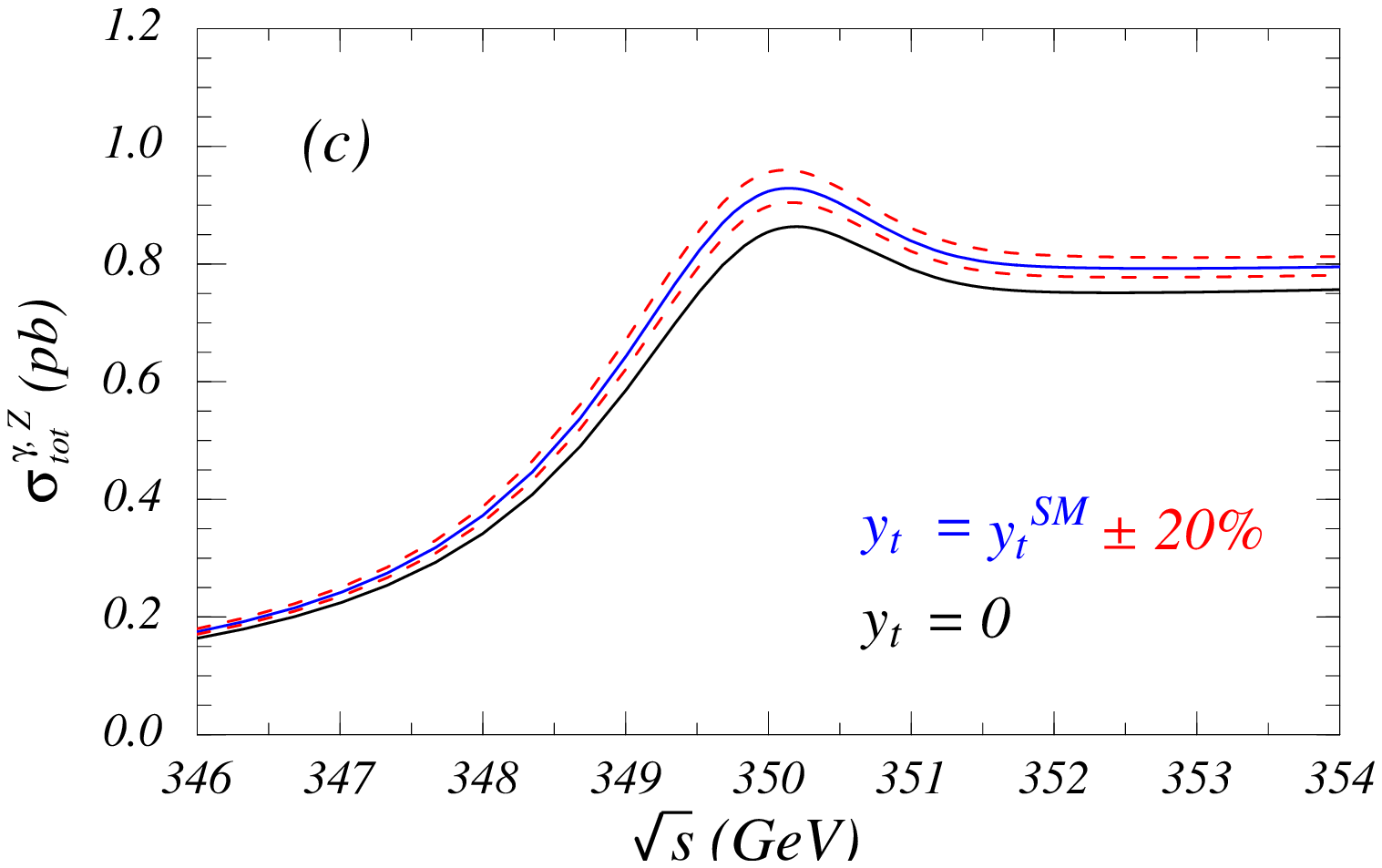}
\vskip 3.cm
 \caption{Variation of the NNLL cross section for a) the value of the
strong coupling, b) the top quark width, and c) the inclusion of a Standard
Model (SM) Higgs boson.  Changes relative to the central value (solid lines) are
shown by dashed red lines. In c) there are two solid lines, the lower black
line is the decoupling limit for the Higgs boson, and the upper blue line is
for a SM Higgs with mass $m_H=115\,{\rm GeV}$.  \label{fig:outlook} }
\end{center}
\end{figure}

Our result for the total cross section confirms the previous result in the fixed
order approach that a top quark mass measurement with theoretical uncertainties
at the level of $100-200$~MeV or better is feasible~\cite{Hoang3}. In principle,
the prospects are now even better, because the small normalization uncertainties
of the renormalization group improved calculation have eliminated the danger
that normalization uncertainties could feed into the top mass measurement
through beam effects caused by initial state radiation, beam energy smearing,
and beam-strahlung.  On the other hand, it is also natural to reconsider the
threshold scan as an instrument to carry out precise measurements of other top
quark parameters such as the strong coupling $\alpha_s$, the top Yukawa coupling
$y_t$ or the total top width $\Gamma_t$ -- ideas that have not been seriously
considered since the large NNLO corrections to the normalization of the cross
section were discovered. In fact, in view of our renormalization group improved
NNLL calculation, older studies assessing how precise these parameters can be
measured (see e.g.~Refs.\ \cite{Fuji1,Comas1}), appear more realistic than
conclusions based on the NNLO predictions.

Here we discuss the sensitivity of the total cross section $\sigma_{\rm
tot}^{\gamma,Z}$ to changes in the input parameters $\alpha_s^{n_l=5}(m_Z)$,
$\Gamma_t$, and $y_t$ (for a light Higgs). Our discussion is restricted to the
Standard Model.  However, we note that low energy supersymmetry does not lead to
considerable changes in the QCD dynamics~\cite{SW}, even though it can lead to
modifications of the value of electroweak parameters that can be measured at the
top threshold such as $\Gamma_t$ and $y_t$.  For the implementation of the Higgs
effects we use the prescription of Ref.\ \cite{Harlander2}. For Higgs mass
$m_H$, the Higgs potential
\begin{eqnarray}
 \tilde V_{tth}(\bmp,\bmq) 
 & = & - \frac{y_t^2}{2}\ \frac{1}{\bmk^2+m_H^2}
\end{eqnarray}
is accounted for exactly through Eq.\ (\ref{CoulombSchroedinger}), and
the one-loop short-distance corrections to the current
$\O{p}{1}$ that depend on $y_t$ are included in the matching
condition $c_1(1)$~\cite{Harlander2}
\begin{eqnarray}
  c_1(1)\ & \to & \ c_1(1) \Bigg[ 1 + \frac{y_t^2}{8\,\pi^2}\,
    \bigg(1 - \frac{r}{3} -\frac{\pi}{\sqrt{r}}  +
    \Big( 1 - \frac{3r}{4} + \frac{r^2}{6} \Big) \ln r \nn\\
 && \qquad\quad + \sqrt{( 4 - r) \,r}\, \Big( \frac{1}{r}- 
    \frac{5}{6} + \frac{r^2}{3} \Big) \,\cos^{-1}\bigg(\frac{\sqrt{r}}{2}
    \bigg)  \Bigg] \,,
\end{eqnarray}
where $r=m_H/m_t$. Assuming that $m_H$ has been measured elsewhere, the question
to address from the top threshold cross-section is how well $y_t$ can be
measured.

In Figs.~\ref{fig:outlook} we have displayed the variation of the NNLL cross
section $\sigma_{\rm tot}^{\gamma,Z}$ as a function of $\sqrt{s}$ for different
choices of the input parameters $\alpha_s(m_Z)$, $\Gamma_t$ and $y_t$.  In
all figures we have chosen $M_t^{\sSS}=175$~GeV, $\Gamma_t=1.43\,{\rm GeV}$,
$\alpha_s(m_Z)=0.118$, $y_t=0$ and $\nu=0.15$, unless stated otherwise.
The upper left panel shows the cross section for $0.116$ (lower dashed red
line), $0.118$ (solid black line), and $0.120$ (upper dashed red line).  At the
peak one finds a $\pm 2.7\%$ variation when varying $\alpha_s(m_Z)$ by $\pm
0.002$.  The upper right panel shows the cross section for $\Gamma_t=1.43$~GeV
(solid black line). The dashed red lines correspond to variations of 
$\Gamma_t$ by $\pm 10\%$, where for a smaller width the peak becomes more
pronounced. At the peak one finds a $(-2.3\%,+2.6\%)$ variation of the cross
section when varying $\Gamma_t$ by $\pm 5\%$.  Finally, the lower panel shows
the cross section for zero Yukawa coupling (solid black line), the Standard
Model value for the Yukawa coupling (solid blue line), and a $\pm 20\%$
variation of the coupling with respect to the SM value (upper/lower dashed red
lines).  The Higgs mass was chosen to be $m_H=115$~GeV.  At the peak one finds a
$(+3.3\%,-2.6\%)$ variation when varying $y_t^{\rm SM}$ by $\pm 20\%$.  For
$m_H=130$ and $150\,{\rm GeV}$ the corresponding variations are
$(+2.9\%,-2.3\%)$ and $(+2.5\%,-2.0\%)$ respectively.

Thus, given the remaining $3\%$ uncertainty from higher order QCD effects,
measurements with uncertainties of $\delta\alpha_s(m_Z)\sim 0.002$,
$\delta\Gamma_t/\Gamma_t\sim 5\%$ and $\delta y_t/y_t\sim 20\%$ appear feasible.
More accurate extractions might be possible if better use is made of the smaller
scale dependence above and below the $\SS$ peak.  A detailed simulation study
based on the NNLL results will be necessary to more realistically assess the
uncertainties in measurements of $\alpha_s$, $\Gamma_t$ and $y_t$. Such a
simulation is beyond the scope of this work.

Finally, we comment on other prospects for measurements of $\alpha_s$,
$\Gamma_t$, and $y_t$.  Extractions of $\alpha_s$ from many processes currently
have a combined $\pm 0.002$--$0.003$ uncertainty~\cite{Bethke,revalphas}, but in
the very long term a $1\%$ uncertainty may be
achievable~\cite{futurealphas,Tesla}.  Measurements of the forward-backward
asymmetry and momentum distribution for top quarks near threshold may help to
reduce the uncertainty in an extraction of $\Gamma_t$~\cite{Fuji1,Comas1}.  The
top Yukawa coupling can also be measured above threshold from $e^+e^-\to t\bar t
h$~\cite{BaerJuste}.  However, at $\sqrt{s}=500\,{\rm GeV}$ the phase space for
this process is restricted, so precise extractions above threshold require a
$0.8$--$1\,{\rm TeV}$ linear collider and high luminosity.

\OMIT{$200$ and $(+1.7\%,-1.4\%)$}

\section{Summary and Outlook}
\label{sectionsummary}

In this work we have studied the impact of the summation of QCD logarithms of
ratios of the scales $m_t$, $m_t v$ and $m_t v^2$ on the total top pair
production cross section in the kinematic regime close to threshold. Using the
recently developed effective field theory vNRQCD\ \cite{Luke1,amis,amis2,amis3}
the logarithms of velocity are summed in the Wilson coefficients of the vNRQCD
operators based on renormalization group equations that are determined from the
anomalous dimensions of the nonrelativistic operators.  All NNLL terms were
accounted for, except for the Wilson coefficient $c_1$ of the dimension three
production current for which only NLL results are completely available\
\cite{amis3}.  Based on partial contributions to the three loop anomalous
dimension that are currently known, the modification of $c_1$ at NNLL order was
estimated to be at the $1\%$ level.  We find that the size of the NNLL
corrections and the variation of the NNLL cross section for different choices of
the scaling parameter $\nu$ is an order of magnitude smaller than the results of
earlier NNLO calculations\ \cite{Hoang3}, where the logarithmic terms were not
summed. We conclude that a conservative estimate of the remaining theoretical
QCD uncertainty of the normalization of the cross section is $\pm 3$\%. The
already excellent prospects of a top quark mass measurement at the level of
$100$--$200$~MeV or better~\cite{Hoang3}, are in principle further improved
because the danger that normalization uncertainties could feed into the top mass
measurement through beam smearing effects is eliminated. Also, measurements of
$\alpha_s$, $\Gamma_t$ and $y_t$ from a threshold scan with theoretical
uncertainties of about $\delta\alpha_s(m_Z)\sim 0.002$,
$\delta\Gamma_t/\Gamma_t\sim 5\%$ and $\delta y_t/y_t\sim 20\%$ appear feasible
at this stage. However, realistic simulation studies are still necessary to
obtain more definite numbers.

At the level of uncertainty of our computation of the total cross section there
are additional subleading effects which should be reconsidered. Many small
effects were previously considered irrelevant for the total cross section due to
the large size of the QCD corrections found at NNLO. Here, let us just mention
electroweak effects. The dominant contribution was included and comes from the
top width, which is the reason that the threshold top cross section is a smooth
lineshape and can be calculated for all center of mass energies.  At subleading
order a number of additional effects need to be taken into account, such as the
so-called non-factorizable corrections\ \cite{nonfactorizable}, electroweak box
and triangle diagrams~\cite{Guth}, $W$-width effects~\cite{KuhnW}, and single- or
non-resonant background processes that lead to the same final state as top pair
production and phase space corrections\ \cite{Hoang2}.  These corrections lead
to effects at the percent level and are expected to be of the same size as the
remaining QCD uncertainties.

\section{Acknowledgement} 
AM and IS are supported in part by the U.S.~Department of Energy under
contract~DOE~DE-FG03-97ER40546. TT is supported in part by BMBF contract 
no.~05-HT1-PAA-4.  IS would like to thank A.~Falk for useful discussion.


\appendix

\section{Constants and Running Couplings} 

\subsection{Formul\ae\ for constants} \label{App_const}

In this appendix expressions for the constants that appear in the effective
Coulomb potential $\tilde V_c({\itbf{p,q}})$ are collected and the two loop
matching for $c_1(1)$ is discussed.  The $\beta$-functions and constant
coefficients which appear in $\tilde V_c({\itbf{p,q}})$ in Eq.~(\ref{VCoulomb})
include
\begin{eqnarray}
 \beta_0 &=& \frac{11}{3}\,C_A - \frac{4}{3}\,T_F\,n_l \,,  \\[2mm]
 \beta_1 &=& \frac{34}{3}\,C_A^2-\frac{20}{3}C_A\,T_F\,n_l- 4\,C_F\,T_F\,n_l
    \,, \nn \\[2mm]
 a_1 &=&  \frac{31}{9}\,C_A - \frac{20}{9}\,T_F\,n_l \,, \nn \\[2mm]
 a_2 &=&  \bigg(\,\frac{4343}{162}+4\,\pi^2-\frac{\pi^4}{4}
   +\frac{22}{3}\,\zeta_3\,\bigg)\,C_A^2 -\bigg(\,\frac{1798}{81}+
   \frac{56}{3}\,\zeta_3\,\bigg)\,C_A\,T_F\,n_l \nn \\[2mm] 
  & & -\bigg(\,\frac{55}{3}-16\,\zeta_3\,\bigg)\,C_F\,T_F\,n_l 
   +\bigg(\,\frac{20}{9}\,T_F\,  n_l\,\bigg)^2 \,, \nn
\end{eqnarray}
where the $a_2$ term was computed in Ref.~\cite{Schroder1}, and $n_l=5$ is the
number of light fermions.

\begin{figure}[t!]
\begin{picture}(180,60)(10,1)
  \put(265,43){$c_1$} \put(305,43){${\cal V}_k$}
  \put(365,43){$c_1$} \put(400,43){${\cal V}_c$} \put(435,43){${\cal V}_c$}
   $\begin{array}{c}
   \centerline{
   \raisebox{15pt}{$\left( \begin{array}{c} \\ \\ \\[2mm] \end{array}\right.
     \!\!\!\!\!\!$}
   \epsfxsize=2.5cm \lower4pt \hbox{\epsfbox{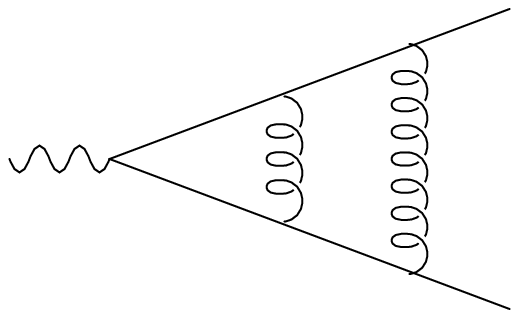}} 
   \raisebox{15pt}{$ + $}
   \epsfxsize=2.5cm \lower4pt \hbox{\epsfbox{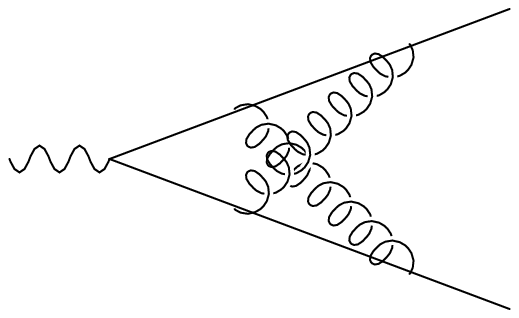}}
   \raisebox{15pt}{$ +\ \ldots $}
   \raisebox{15pt}{$\!\!\!\! \left. \begin{array}{c} \\ \\ \\[2mm] \end{array}
     \right)$}
   \raisebox{15pt}{$-$} 
   \raisebox{15pt}{$\left( \begin{array}{c} \\ \\[2mm] \end{array}\right.
      \!\!\!\!$}
   \epsfxsize=2.5cm \raise4pt \hbox{\epsfbox{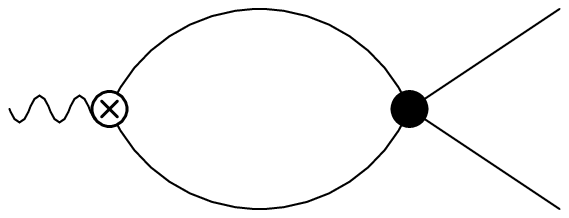}} \quad
   \raisebox{15pt}{$ + $}
   \epsfxsize=3.5cm \raise4pt \hbox{\epsfbox{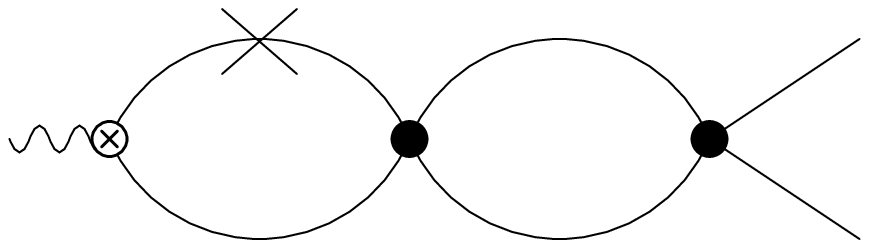}} 
   } \\[5mm]
   \put(53,33){$c_1$} \put(85,33){${\cal V}_{2,s,r}$} \put(125,33){${\cal V}_c$}
  \put(182,33){$c_1$} \put(215,33){${\cal V}_c$} \put(245,33){${\cal V}_{2,s,r}$}
  \put(312,33){$c_2$} \put(345,33){${\cal V}_c$} \put(380,33){${\cal V}_c$}
   \centerline{
   \raisebox{15pt}{$ + $}
   \epsfxsize=3.5cm \raise4pt \hbox{\epsfbox{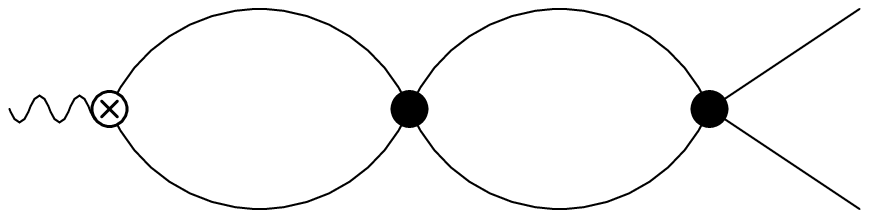}} \quad
   \raisebox{15pt}{$ + $}
   \epsfxsize=3.5cm \raise4pt \hbox{\epsfbox{figures/c11b.eps}} \quad
   \raisebox{15pt}{$ + $}
   \epsfxsize=3.5cm \raise4pt \hbox{\epsfbox{figures/c11b.eps}} \quad
   \raisebox{15pt}{$\left. \begin{array}{c} \\ \\[2mm] \end{array}\right)$}
   } 
   \end{array} $
\end{picture}
\vskip 2.cm 
\caption{Difference of full QCD~\cite{Melnikov1} and order $\alpha_s^2 v^0$ EFT
graphs which gives the two loop matching for $c_1(1)$. The $\times$ denotes an
insertion of the ${\bmp}^4/(8m^3)$ operator, and graphs with this insertion on
a different propagator are understood.\label{fig_c1match}}
\end{figure}

For the matching onto the coefficient $c_1(1)$ of the non-relativistic current
$\O{p}{1}$ in Eq.~(\ref{Ov}), the one-loop contribution is well known. The
two-loop computation involves computing the difference of graphs in full QCD and
the effective theory as shown in Fig.~\ref{fig_c1match}. We find
\begin{eqnarray} \label{c1match}
c_1(1) & = & 1- \frac{2 C_F}{\pi}\: {\alpha_s(m)} +   
  \alpha_s^2(m) \bigg[C_F^2\Big(\frac{\ln 2}{12}-\frac{25}{24}
  -\frac{2}{\pi^2}\Big) + C_A C_F\Big(\ln 2-1\Big) + \frac{\kappa}{2} \bigg] 
  \,, \\[3mm]
\kappa & = & C_F^2 \bigg[ \frac{1}{\pi^2}\Big(\frac{39}{4}-\zeta_3\Big) +
  \frac{4}{3}\ln 2 - \frac{35}{18} \bigg] - C_A C_F \bigg[ \frac{1}{\pi^2} 
  \Big( \frac{151}{36} + \frac{13}{2} \zeta_3 \Big) +
  \frac{8}{3} \ln 2 - \frac{179}{72} \bigg] \nonumber\\[2mm]
 & & + C_F T_F \bigg[ \frac{4}{9} \Big( \frac{11}{\pi^2} - 1 \Big) \bigg] +
  C_F T_F n_l \bigg[ \frac{11}{9 \pi^2} \bigg] 
  \,.\nn
\end{eqnarray}
At order $\alpha_s(m)^2$ this matching result is scheme dependent and depends
both on the subtraction scheme as well as on the definition of operators in the
effective theory.  For this reason the result in Eq.~(\ref{c1match}) differs
from the earliest results in Refs.~\cite{Czarnecki1,Beneke4} for $c_1(1)$.  In
Refs.~\cite{Czarnecki1,Beneke4} the potential operators in Eq.~(\ref{Vkdr}) were
defined with dependence on $d=4-2\epsilon$ in such a way that only the hard part
of the two loop graphs in the threshold expansion contributed.  The result in
Eq.~(\ref{c1match}) uses the $\overline{\rm MS}$ scheme with all effective
theory operators defined on-shell in $d=4$ space-time dimensions. The full
theory result was extracted from Ref.~\cite{Czarnecki1}, while the EFT graphs
were computed explicitly.  We have obtained the same result by using the
``direct matching''of Ref.~\cite{Hoang4}, where one matches results for cross
sections in the full and effective theories rather than matching the Green
functions.
%
%

\subsection{Formul\ae\ for running couplings} \label{App_run}

In this section the formul\ae\ for the Wilson coefficients of operators needed
for our NNLL cross section are given.  The Wilson coefficients for the potential
operators in Eqs.~(\ref{VCoulomb}) and (\ref{Vkdr}) were computed in
Refs.~\cite{amis,amis3,hms1}, and due to mixing depend on the one-loop running
of the HQET $1/m^2$ terms computed in Ref.~\cite{bauer}. For completeness we
summarize the results for the color singlet coefficients:
\begin{eqnarray}  \label{pv2}
 {\cal V}_c^{(s)}(\nu) &=& -4 \pi C_F\, z\, \alpha_s(m)  + 
   \frac{8\pi C_F C_A^3}{3\beta_0}\, \bigg[\frac{11}{4}-2z 
   -\frac{z^2}{2}-\frac{z^3}{4}+4 \ln(w) \bigg]\, \alpha_s^3(m) \,,\nn \\[2mm]
 {\cal V}_k^{(s)}(\nu)  &=& C_F\Big(\frac{C_F}{2}-C_A\Big) z^2\, \alpha^2(m) 
   +\frac{8 C_F C_A (C_A+2C_F)}{3\beta_0}  \bigg[4-3 z-z^2 
   + 6 \ln(w) \bigg]\, \alpha_s^2(m) \,,\nn \\[2mm]
 {\cal V}_2^{(s)}(\nu) &=& C_F\pi \Bigg\{
 \frac{235 C_A^2-716 C_F C_A+99\beta_0 C_A + 96 \beta_0 C_F}{39\beta_0 C_A} 
 [z-1] \nn\\[2mm]
 &-& \frac{(\beta_0-5 C_A)}{(\beta_0-2 C_A)}\Big[ z^{(1-2C_A/\beta_0)}-1 \Big]
   -\frac{8(3\beta_0-11 C_A)(5 C_A + 8 C_F)}{13(6\beta_0-13 C_A)C_A}
  \Big[ z^{(1-13C_A/(6\beta_0))}-1 \Big]\nn\\[2mm]
 &-& \frac{32(C_A-2C_F)}{3\beta_0} \ln(w) \Bigg\} \alpha_s(m)  \,, \nn\\[2mm]
 {\cal V}_s^{(s)}(\nu) &=& \frac{-2 \pi C_F}{(2 C_A-\beta_0) } 
  \bigg[ C_A + \frac{1}{3} ( 2\beta_0 - 7 C_A) \ z^{(1-2 C_A/\beta_0)} \bigg]
  \, \alpha_s(m) \,, \nn\\[2mm]
 {\cal V}_r^{(s)}(\nu) &=& -4\pi C_F z\, \alpha_s(m) + \frac{32\pi C_F C_A}
  {3\beta_0} \Big[ 1 - z \Big]\,\alpha_s(m)  + \frac{64\pi C_F C_A}{3\beta_0}\,
   \ln( w )\, \alpha_s(m) \,,   
\end{eqnarray}
where
\begin{eqnarray} \label{zw}
  z=\frac{ \alpha_s(m\nu) }{ \alpha_s(m) }\,,\qquad\quad
  w=\frac{ \alpha_s(m\nu^2) }{ \alpha_s(m\nu) } \,.
\end{eqnarray}

%
\begin{figure}[t!]
\begin{picture}(380,60)(-7,1)
  \put(30,-3){$c_1$} \put(90,-4){${\cal V}_c$}
  \put(170,-3){$c_1$} \put(210,-3){${\cal V}_c$} \put(252,-3){${\cal V}_c$}
  \put(333,-3){$c_1$} \put(368,-3){$\delta{\cal V}_{2,r}$} 
       \put(412,-3){${\cal V}_c$}
   \epsfxsize=3.9cm \lower4pt \hbox{\epsfbox{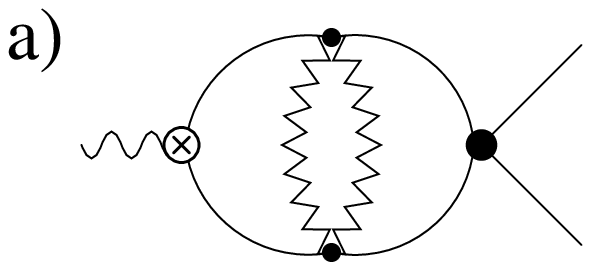}} \qquad\quad
   \epsfxsize=4.3cm \raise4pt \hbox{\epsfbox{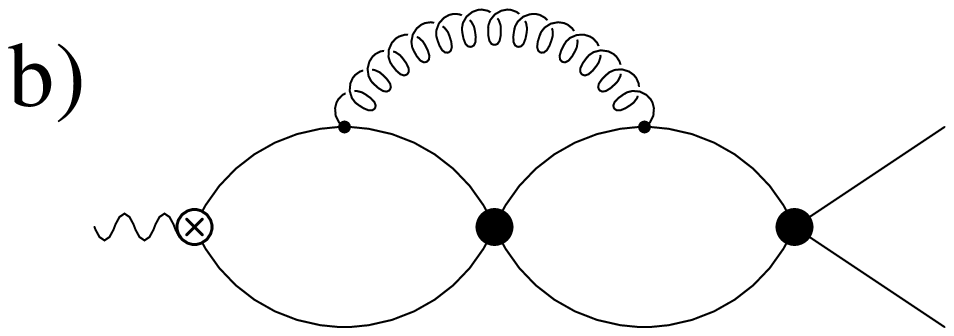}} \qquad\quad
   \epsfxsize=4.3cm \raise4pt \hbox{\epsfbox{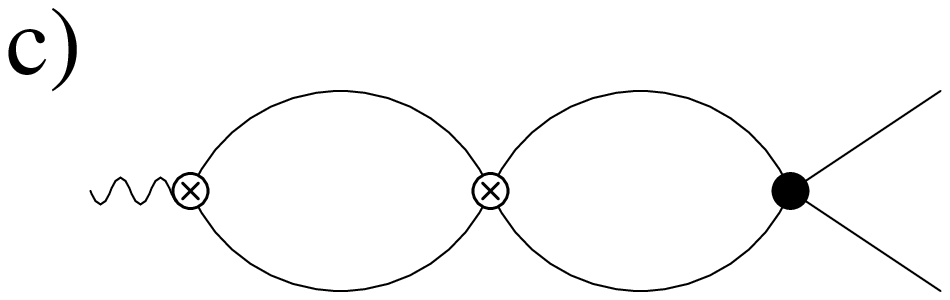}} 
\end{picture}
\vskip 0.8cm 
\caption{Examples of the type of graphs that contribute to the three-loop
anomalous dimension for $c_1$. $\delta {\cal V}_{2,r}$ are one loop
counterterms. \label{fig:c1run}}
\end{figure}
Next consider the Wilson coefficients of the currents in Eqs.~(\ref{Ov}) and
(\ref{Oa}).  The LL solutions are $c_1(\nu)=c_3(\nu)=1$, while the solution for
$c_2(\nu)$ is given in Eq.~(\ref{c2ll}).  The NLL solution for $c_1(\nu)$ is 
obtained by integrating the two-loop anomalous dimension~\cite{Luke1}
\begin{eqnarray} \label{adc1}
 \nu {\partial  \over \partial\nu} \ln[c_1(\nu)] 
 &=& -{{\cal V}_c^{(s)}(\nu)
  \over 16\pi^2} \left( { {\cal V}_c^{(s)}(\nu) \over 4 }
  +{\cal V}_2^{(s)}(\nu)+{\cal V}_r^{(s)}(\nu)
  + 2 {\cal V}_s^{(s)}(\nu)  \right) +
  { {\cal V}_{k}^{(s)}(\nu) \over 2} \,, 
\end{eqnarray}
with one-loop matching. The analytic solution of Eq.~(\ref{adc1}) can be 
found in Ref.~\cite{amis3}.

At NNLL we use the two-loop matching in Eq.~(\ref{c1match}), however the
complete three-loop anomalous dimension is currently unknown.  Part of the
unknown contributions are due to the mixing in Eq.~(\ref{adc1}). Solving for the
full NNLL $c_1(\nu)$ requires the NLL values of ${\cal V}_c(\nu)$, ${\cal
V}_2(\nu)$, ${\cal V}_r(\nu)$ and ${\cal V}_s(\nu)$, and the NNLL value of
${\cal V}_k(\nu)$, which in turn require knowing the running of $1/m^2$
operators in the HQET Lagrangian at two-loops. Note that at one-loop, the
calculation of the full mixing matrix in Ref.~\cite{bauer} already involved 75
diagrams.  In addition there are direct contributions to the three loop
anomalous dimension from three-loop graphs with divergences which require a
$c_1$ counterterm.  By power counting we find that there are no contributions
from purely potential three loop diagrams. However, diagrams with soft and
ultrasoft gluons such as those shown in Fig.~\ref{fig:c1run} may
contribute. Graphs with ultrasoft gluons are expected to generate an anomalous
dimension of the form
\begin{eqnarray} \label{c1lambda}
 \gamma_{\rm us} &=& \lambda\ {\alpha_s(m\nu^2) [{\cal V}_c^{(s)}(\nu)]^2 \over 
   64\pi^3}
\end{eqnarray}
where $\lambda$ is (currently) an unknown number.  A complete calculation of
$\lambda$ may not be that difficult, and is of interest since the graphs with
soft gluons give contributions which are parametrically smaller by a factor
$\alpha_s(m\nu)/\alpha_s(m\nu^2)$.

Interestingly, we note that the cancellation of subdivergences $\ln(p)/\epsilon$
and $\ln(E)/\epsilon$ between Fig.~\ref{fig:c1run}b and the counterterm graph in
Fig.~\ref{fig:c1run}c {\em requires} the velocity renormalization group relation
$\mu_U=\mu_S^2/m$ for consistency. In Coulomb gauge
\begin{eqnarray} \label{c1lamest}
 {\rm Fig.~\ref{fig:c1run}b} &=& {\alpha_s(m\nu^2) [{\cal V}_c(\nu)]^2 \over
 192\pi^3} \Big(C_F-\frac{C_A}{2}\Big) \Bigg[ \frac{1}{4\epsilon^2} + 
 \frac{1}{4\epsilon} \ln\Big(\frac{\mu_U^2}{E^2}\Big) + 
 \frac{1}{2\epsilon}\ln\Big(\frac{\mu_S^2}{p^2}\Big) + \frac{\#}{\epsilon} + 
 {\rm finite} \Bigg] \,,\nn \\
 {\rm Fig.~\ref{fig:c1run}c} &=& -{\alpha_s(m\nu^2) [{\cal V}_c(\nu)]^2 \over
 192\pi^3} \Big(C_F-\frac{C_A}{2}\Big) \Bigg[ \frac{1}{2\epsilon^2} + 
 \frac{1}{\epsilon}\ln\Big(\frac{\mu_S^2}{p^2}\Big) + \frac{\#}{\epsilon} + 
 {\rm finite} \Bigg] \,. 
\end{eqnarray}
It is easy to see that the subdivergences of the form $\ln(\cdots)/\epsilon$
cancel completely {\em only} after using the equations of motion $E=p^2/m$ and
$\mu_U=\mu_S^2/m$. The difference of $\#/\epsilon$ terms in Eq.~(\ref{c1lamest})
will contribute to Eq.~(\ref{c1lambda}). If only a single $\mu$ were used then
one would be left with a $\ln(\mu/m)$ term contributing to the anomalous
dimension which is obviously wrong.

To numerically estimate the effect of the three loop anomalous dimension on
$c_1(\nu)$ we consider several types of contributions:
\begin{enumerate}

\item At NLL the running of ${\cal V}_c^{(s)}(\nu)$ is determined by the
two-loop $\beta$-function, so a subset of terms in the anomalous dimension are
determined by increasing the order of the running used for $\alpha_s$. Using
one-loop running for $z$ and $w$ in Eq.~(\ref{zw}) gives $c_1^{\rm
NNLL}(\nu=0.15)=0.8970$, whereas two-loop running gives $c_1^{\rm
NNLL}(\nu=0.15)=0.8941$.

\item Another estimate comes from partial knowledge of the NLL results for
${\cal V}^{(s)}_{2,s}(\nu)$ and the NNLL result for ${\cal V}^{(s)}_k(\nu)$
which enter in Eq.~(\ref{adc1}). The matching for these potentials is known,
however the complete anomalous dimensions are not. The one-loop matching values
are~\cite{PSmatch,amis2} ${\cal V}_2^{(s)}(1)= 241/135\, \alpha_s^2(m)$ and
${\cal V}_s^{(s)}(1)= 112/27\, \alpha_s^2(m)$. The recently computed two-loop
matching~\cite{KPSS} for the $1/|{\bf k}|$ potential gives ${\cal
V}_k(1)=-(289\!+\!4896\ln 2)/(324\pi)\,\alpha_s^3(m)$.  Furthermore, the QED
anomalous dimensions for these potentials are known~\cite{amis4}, and directly
give a majority of the purely potential contributions to the QCD anomalous
dimensions (in the notation of Ref.~\cite{amis4} we include the terms
$\rho_{ccc}$, $\rho_{cc2}$, $\rho_{c22}$, $\rho_{ck}$, and $\rho_{k2}$). Using
these results as approximations for ${\cal V}_{2,s}^{(s)\,\rm NLL}(\nu)$ and
${\cal V}_{k}^{(s)\,\rm NNLL}(\nu)$, and then including these contributions in
Eq.~(\ref{adc1}) gives $c_1^{\rm NNLL}(\nu=0.15)=0.8991$. The result without the
two-loop matching from Ref.~\cite{KPSS} is $c_1^{\rm NNLL}(\nu=0.15)=0.8915$.

\item Finally, adding a contribution of the form in Eq.~(\ref{c1lambda}) with
$\lambda=\pm 5$ to account for the possibility of a large coefficient gives the
range $c_1^{\rm NNLL}(\nu=0.15)=0.8903$ to $0.8969$ (using three loop running
for $\alpha_s$).

\end{enumerate}
Adding the magnitude of these three anomalous dimension contributions linearly,
we find a $1.2\%$ change in $c_1(\nu=0.15)$ relative to the NNLL value which
includes only the two-loop matching for $c_1$.  We therefore do not expect that
the missing three loop anomalous dimension for $c_1$ will shift the value of
$R^v$ by more than an amount at the $2\%$ level.

\end{document}